\documentclass[11pt,letterpaper]{article}
\pdfoutput=1
\usepackage{jcappub}

\usepackage{cancel}
\usepackage{graphicx}
\usepackage{dcolumn}
\usepackage{bm}
\usepackage[usenames,dvipsnames]{xcolor}
\usepackage{soul}
\usepackage{subcaption}
\usepackage{enumitem}
\usepackage{xspace}
\usepackage{booktabs}
\usepackage{lipsum}
\usepackage{siunitx} 
\usepackage[ISO]{diffcoeff}
\usepackage{subcaption}
\usepackage{ragged2e}
\usepackage{mathrsfs}
\usepackage{hyperref}
\usepackage[capitalise]{cleveref}
\usepackage[normalem]{ulem}
\usepackage{amsmath}
\usepackage{wrapfig}
\usepackage{esvect} 
\usepackage{titlesec}
\usepackage{titletoc}
\usepackage{chngcntr}
\usepackage{fontawesome5}
\usepackage{tikz}
\usepackage{mleftright}
\usepackage[nodayofweek]{datetime}
\usepackage{comment}

\newcommand{\Evec}{{\bm E}}
\newcommand{\Bvec}{{\bm B}}

\newcommand{\Xvec}{{\bm{X}}}

\newcommand{\MYhref}[3][blue]{\href{#2}{\color{#1}{#3}}}%

\newcommand\funop[1]{\mathop{{}#1}}
\newcommand{\dd}{\mathop{}\!\mathrm{d}}
\newcommand{\mrm}[1]{\mathrm{#1}}
\newcommand{\usim}{\mathord{\sim}}
\newcommand{\ulesssim}{\mathord{\lesssim\,}}

\newcommand{\bvec}[1]{\boldsymbol{#1}}

\DeclareCaptionJustification{justified}{\justifying}
\def\arraystretch{2}\tabcolsep=10pt

\definecolor{firebrick}{HTML}{B22222}
\definecolor{mycolor}{HTML}{205584}
\definecolor{orcid-green}{RGB} {166, 206, 57}

\hypersetup{
    colorlinks=true,       
    linkcolor=mycolor,          
    citecolor=mycolor,        
    filecolor=mycolor,      
    urlcolor=mycolor          
}

\title{\huge Constraining Dark Photon Dark Matter with Radio Silence from Soliton Mergers around Supermassive Black Holes}

\author[a,b]{Dorian W.~P.~Amaral {\MYhref[orcid-green]{https://orcid.org/0000-0002-1414-932X}{\faOrcid}},}
\emailAdd{damaral@ifae.es}

\author[c]{Enrico D.~Schiappacasse {\MYhref[orcid-green]{https://orcid.org/0000-0002-6136-1358}{\faOrcid}},}
\emailAdd{enrico.schiappacasse@uss.cl}

\author[d]{Hong-Yi Zhang {\MYhref[orcid-green]{https://orcid.org/0000-0003-4176-5707}{\faOrcid}}}
\emailAdd{hongyi18@sjtu.edu.cn}

\affiliation[a]{Department of Physics and Astronomy, Rice University, 6100 Main Street, Houston, TX 77005, U.S.A.}

\affiliation[b]{Institut de F\`{ı}sica d’Altes Energies (IFAE), The Barcelona Institute of Science and Technology,
Campus UAB, 08193 Bellaterra (Barcelona), Spain}

\affiliation[c]{Facultad de Ingenier\'ia, Universidad San Sebasti\'an, Bellavista 7, Santiago 8420524, Chile}

\affiliation[d]{Tsung-Dao Lee Institute \& School of Physics and Astronomy, Shanghai Jiao Tong University, Shanghai 201210, China}

\date{\today}

\abstract{We place the first constraints on the dark matter fraction contained within dark photon solitons using the absence of their predicted radio-frequency signatures, or \textit{radio silence}, following mergers around supermassive black holes.  In these dense environments, spiky dark matter density profiles can form that enhance the soliton merger rate. We present a novel estimate of this rate by incorporating both the steepened dark matter profile and the soliton velocity dispersion via the Jeans equation. For galaxies with an initial profile $\rho_\mathrm{DM} \propto r^{-1}$, we find the total merger rate across redshifts $0 \leq z \leq 4$ to be $\Gamma_{\text{merg}}^{\text{TOTAL}} \lesssim 10^{-7}f^2_{\text{DM}}\,\text{Mpc}^{-3}\,\text{day}^{-1}$, where $f_\mathrm{DM}$ is the solitonic fraction of dark matter. This enhanced rate leads to more major merger events in which the generated soliton has a mass exceeding a critical threshold, leading to its decay via the parametric resonance phenomenon that produces brief, narrowband, and energetic radio bursts detectable by fast radio burst surveys. Comparing our predictions with the non-observation of such events, we already obtain $f_\mathrm{DM} \lesssim 10^{-1}$ from the first fast radio burst study. This constraint is strengthened to $f_\mathrm{DM} \lesssim 10^{-2}$ from the Parkes HTRU survey, with CHIME projected to tighten this to $f_\mathrm{DM} \lesssim 10^{-3}$. For larger $f_\mathrm{DM}$, we instead constrain the effective coupling strength between the dark and visible sectors to lie outside $10^{-18}\,\mathrm{GeV^{-1}} \lesssim g \lesssim 10^{-8}\,\mathrm{GeV^{-1}}$ for dark photon masses in the range $10^{-6}\,\mathrm{eV} \lesssim m \lesssim 10^{-4}\,\mathrm{eV}$. Our results establish astrophysical transients as powerful probes of dark sectors, opening a window onto the detectability of ultralight vector fields.  }

\begin{document}

\maketitle

\tableofcontents

\section{Introduction}
\label{sec:intro}

Dark matter (DM)  dominates the non-relativistic matter content of our universe. Despite this, we understand remarkably little about its fundamental nature. While we know that it must interact gravitationally, important properties such as its mass, spin, and other possible interactions remain a mystery ~\cite{Bertone:2016nfn,Freese:2017idy,Cirelli:2024ssz}. 
Astrophysical observations suggest that its mass can lie anywhere in the range of $10^{-19}\,\mrm{eV}$ to a few solar masses, spanning almost 90 orders of magnitude~\cite{Brandt:2016aco,Dalal:2022rmp, Amin:2022nlh}.
The lower end of this mass window, $\ulesssim 10\,\mathrm{eV}$, defines the \textit{ultralight} regime for DM, which has been gaining considerable attention~\cite{Essig:2013lka,Antypas:2022asj}.

In the ultralight regime, DM particles must be bosonic to reconcile the observed DM density, leading to wavelike behavior~\cite{Tremaine:1979we}. Popular ultralight candidates include the QCD axion~\cite{Weinberg:1977ma,Wilczek:1977pj,DiLuzio:2020wdo}, axionlike particles as well as other scalars~\cite{Arvanitaki:2009fg, Ringwald:2014vqa, Ferreira:2020fam}, and---of relevance to us---vector particles~\cite{Jaeckel:2012mjv,Fabbrichesi:2020wbt,Caputo:2021eaa}.
%
Also known as dark photons, vector particles are spin-$1$ bosons that can stem from the same kind of symmetry as the Standard Model photon. Several early-universe production mechanisms have been proposed that generate DM as ultralight vector bosons. These include those related to inflation~\cite{Graham:2015rva,Bastero-Gil:2018uel, Kolb:2020fwh, Ozsoy:2023gnl}, axion oscillations~\cite{Agrawal:2018vin, Co:2018lka}, a dark Higgs~\cite{Dror:2018pdh}, Abelian-Higgs cosmic strings~\cite{Long:2019lwl}, dilatonlike scalar field oscillations~\cite{Adshead:2023qiw}, and axion oscillons~\cite{Zhang:2025lwr}.

Ultralight vector bosons can condense into astrophysical compact objects known as solitons---cosmologically long-lived, spatially localized, and coherently oscillating Bose-Einstein condensates~\cite{Jain:2021pnk, Zhang:2021xxa, Zhang:2024bjo}. Solitons can form via several mechanisms; for example, inflationary isocurvature fluctuations of the dark photon field can gravitationally collapse around the matter-radiation equality to form solitonic structures~\cite{Gorghetto:2022sue}, or dark photons within virialized DM halos can gravitationally condense in the kinetic regime~\cite{Amin:2022pzv, Jain:2022agt, Jain:2023ojg,Jain:2023qty, Chen:2024pyr}. Crucially, if the dark photon field couples to electromagnetism, solitons can undergo a process known as \textit{parametric resonance} during which they produce bursts of radio-frequency radiation detectable by radio telescopes. This makes solitons astrophysical laboratories within the framework of DM indirect searches~\cite{Hertzberg:2020dbk, Du:2023jxh, Escudero:2023vgv, Amin:2023imi,Chen:2024vgh, Amin:2020vja}.

Parametric resonance is a ubiquitous phenomenon throughout physics, occurring when the parameters of an oscillating system are themselves periodic. It features, for instance, in classical mechanics~\cite{seyranian2004swing}, nonlinear optics~\cite{2012OExpr..2025096A, 2021QuEle..51..692M, PhysRevA.87.063848}, electrical systems~\cite{Biryuk2019}, and early-universe cosmology~\cite{Kofman:1994rk, 1995PhRvD..51.5438S, Amin:2014eta}. When solitons undergo parametric resonance, the coherent oscillations of the dark photon field provide a time-periodic background that resonantly amplifies the fluctuations in the electromagnetic field. This culminates in a highly efficient, exponentially enhanced transfer of energy into the visible photon modes. The emitted burst of radiation is narrowband and sharply peaked in the radio-frequency spectrum if the constituent bosons have masses approximately in the range \(10^{-6}\,\text{eV}\) to \(10^{-4}\,\text{eV}\). These bursts can carry energies that significantly exceed those of already observed fast radio bursts (FRBs)~\cite{Petroff:2016tcr,Katz:2018xiu,Petroff:2019tty}. Thus, this mechanism provides a powerful observational channel through which spin-$1$ DM can be detected via transient radio signals.

However, not all dark photon solitons are able to undergo parametric resonance. The resonance is only triggered when the coupling between spin-$1$ particles and electromagnetism is large enough or, equivalently, when the mass of the soliton surpasses a critical threshold \cite{Amin:2023imi}. All supercritical solitons will parametrically resonate, removing mass from these compact objects as dark photons are converted into electromagnetic radiation. This process continues until their masses fall below the critical mass, at which point the parametric resonance phenomenon is once again switched off. Therefore, we expect any remaining vector solitons within DM galactic halos to be subcritical, possessing masses just below the critical mass.

While these subcritical solitons are forbidden from resonating, the merger of such solitons can create a supercritical soliton, reactivating the resonance process. This idea was first proposed for scalar solitons in Ref.~\cite{Hertzberg:2018zte}, developed in Refs.~\cite{Hertzberg:2020dbk, Du:2023jxh}, and extended to spin-$2$ particles in Ref.~\cite{Schiappacasse:2025mao}. As demonstrated by several numerical simulations, collisions between solitons lead to mergers if the initial total energy of the colliding system is negative, corresponding to a bound system~\cite{Hertzberg:2018zte, PhysRevD.74.063504, PhysRevD.83.103513, PhysRevA.54.2185, PhysRevLett.78.4143}. Consequently, when calculating galactic soliton merger rates, we require an accurate assessment of soliton kinetic energies, which can be achieved with a thorough calculation of the relative velocity distribution functions between solitons. 

Generally, solitons with typical dispersion velocities of \(\usim \mathcal O(100) \,\mrm{km\,s^{-1}}\), such as those expected in the dark halos surrounding elliptical and spiral galaxies\,\cite{2001AJ....121.1936G, Salucci:2007tm, Bose:2020xug}, carry too much energy, leading to a negligible merger rate as solitons merely pass through one another. For galaxies similar to the Milky Way, the typical merger rate is at most \( \mathcal{O}(10^{-10}) \, \mrm{day^{-1}\,galaxy^{-1}} \) \cite{Schiappacasse:2025mao, Hertzberg:2020dbk}. However, since the merger rate is proportional to the square of the dark soliton number density, we anticipate a significant enhancement of this rate under favorable astrophysical scenarios. One such scenario involves the presence of supermassive black holes (SMBHs) at galactic centers, and it is well established that nearly all large galaxies host these astrophysical objects~\cite{1995ARA&A..33..581K, 2013ARA&A..51..511K}. The growth of SMBHs can lead to a highly concentrated, or `spiky' DM profile in the central regions of galaxies~\cite{Gondolo:1999ef, Bertone:2002je, Zhang:2025mdl}, as indicated by observations of the SMBH binary OJ 287 \cite{Chan:2024yht}. These spiky DM profiles have also been extensively used to place significant constraints on the DM annihilation rate \,\cite{2000PhLB..494..181G, Bertone:2002je, Fields:2014pia, Lacroix:2016qpq, 2022SCPMA..6569512L, Balaji:2023hmy}.
The increased DM density in these regions results in sizable rates, leading to more supercritical solitons emitting electromagnetic signals via parametric resonance.

The signals produced by the decay of supercritical solitons can be readily searched for by radio telescopes. Since the produced radiation is brief, sharp, and spectrally confined, it is an ideal target for transient radio surveys searching for FRBs. These include the Parkes radio telescope~\cite{Lorimer:2007qn,2016MNRAS.455.2207R}, the Canadian Hydrogen Intensity Mapping Experiment (CHIME) Pathfinder~\cite{CHIMEScientific:2017php}, the Australian Square Kilometre Array Pathfinder (ASKAP)~\cite{2018Natur.562..386S}, and the Greenbank Telescope (GBT)~\cite{Gajjar:2018bth}. These telescopes have already scanned large regions of the sky with enough sensitivity to detect these bursts of radiation---yet, no such burst have been observed. Using the non-observation of these radio-frequency signatures, or \textit{radio silence}, we can  constrain the rate at which spin-$1$ solitons merge throughout the universe. Since this merger rate depends on the abundance of vector solitons, we may in turn constrain the fraction of DM that they can compose as a macroscopic candidate or, for larger DM fractions, the coupling strength between the dark photon and photon modes over a range of dark photon masses controlled by the operable frequency range of a telescope. 

\begin{figure*}[!t]
    \centering
    \includegraphics[width=0.93\textwidth]{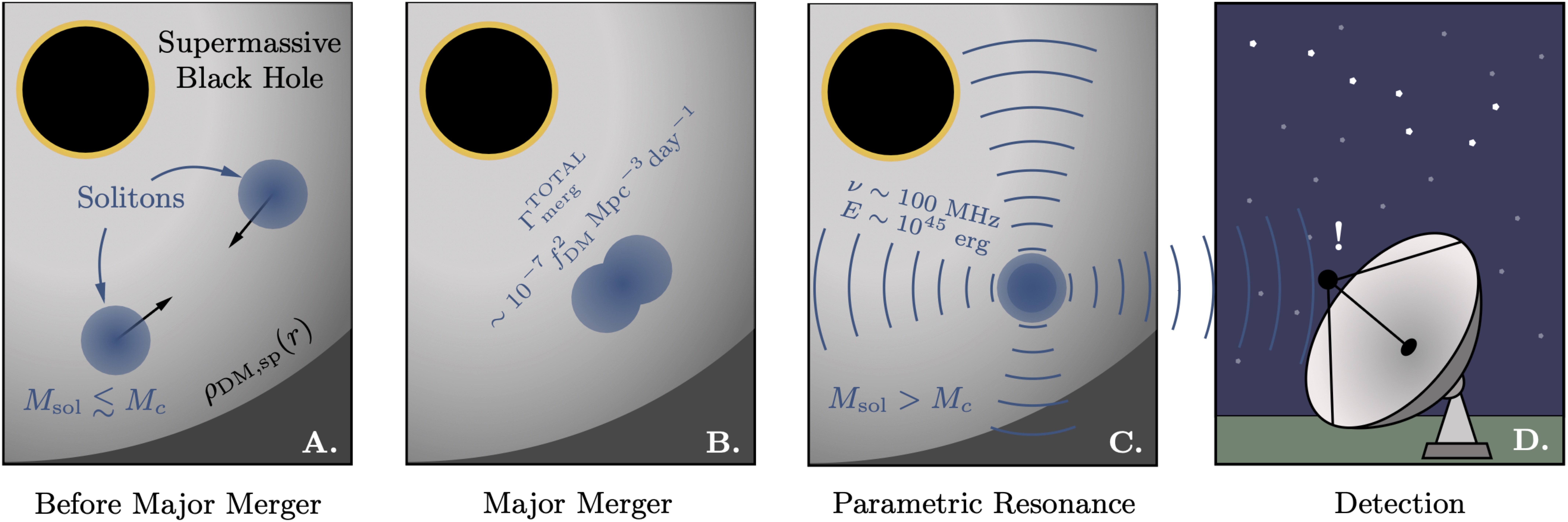}
    \caption{\justifying The merging of two solitons in the spiky dark matter halo around a supermassive black hole, producing a detectable burst of radio-frequency radiation. \textbf{A.} Before a major merger event. Two subcritical solitons with masses around or below the critical mass, $M_\mrm{sol} \lesssim M_\mrm{c}$, travel towards one another.  This occurs within the spiky dark matter halo formed around a supermassive black hole and has a radial density profile $\rho_\mrm{DM,sp}(r)$. \textbf{B.} The major merger event, which occurs at a total rate of $\Gamma_\mrm{merg}^\mrm{TOTAL}$ across redshifts $0 \leq z \leq 4$. \textbf{C.} The production of electromagnetic radiation via parametric resonance due to the decay of the supercritical remnant soliton with mass $M_\mrm{sol} > M_\mrm{c}$. The radiation can have a frequency in the radio frequency range, $\nu \sim 100\,\mrm{MHz}$, and energy $E \sim 10^{45}\,\mrm{erg}$. \textbf{D.} The produced radiation travels to the Earth, where it is detectable by radio telescopes.}
    \label{fig:story}
\end{figure*}

In this article, we compute the first bound on the fraction of DM that can reside in dark photon solitons using the predicted radiation output from their mergers. We adopt a phenomenological approach, avoiding assumptions regarding a specific formation mechanism for dark photon solitons. We begin by providing a brief review on dark photon solitons in \cref{sec:soliton_mergers}. We then carry out a novel and robust estimate of the vector soliton merger rate in the inner regions of galaxies accounting for the spiky DM profiles induced by the presence of central SMBHs.
Finally in \cref{sec:param-res}, we combine our merger rate calculation with radio telescope exposure data and our signal predictions from the parametric resonance phenomenon to draw inferences on dark matter properties. We place upper bounds on the DM fraction in vector solitons and the strength of the dark photon-photon coupling using results from several FRB telescopes and surveys: The Parkes radio telescope~\cite{Lorimer:2007qn,2016MNRAS.455.2207R}, CHIME Pathfinder~\cite{CHIMEScientific:2017php}, ASKAP~\cite{2018Natur.562..386S}, and a high-exposure realization of GBT~\cite{Gajjar:2018bth}. 
In \cref{fig:story}, we schematically summarize the entire process---from the pre-merger dynamics to the potential detection of a radio burst. Our results illustrate how concrete astrophysical predictions can shape and constrain the dynamics, interactions, and viability of ultralight vector field models.\footnote{Throughout this work, we make use of natural units, such that $\hbar = c = 1$.}

\section{Soliton Mergers}
\label{sec:soliton_mergers}

\subsection{Dark Photon Solitons}
\label{sec:darkphotonsolitons}

A population of dark photon particles can cluster into solitons, which achieve stability through a balance of kinetic pressure and the self-gravity of the dark photons. Although each dark photon particle is extremely light, the resulting solitons can acquire macroscopic masses and can themselves constitute a viable form of macroscopic DM. Their properties can be constrained through astrophysical observations, such as gravitational microlensing~\cite{Croon:2020wpr, Croon:2020ouk, Fujikura:2021omw}.

In the non-relativistic regime, the equations of motion of the dark photon field admit the following base solutions of dark photon solitons \cite{Jain:2021pnk, Zhang:2021xxa}
\begin{equation}
\begin{split}
\boldsymbol X^{(0)}(t,\boldsymbol x) &= \funop{X(r)} \cos(\omega_0 t) \,\hat{\boldsymbol z} ~,\\
\boldsymbol X^{(\pm 1)}(t,\boldsymbol x) &= \frac{1}{\sqrt{2}} \funop{X(r)} [ \cos(\omega_0 t) \,\hat{\boldsymbol x} \pm \sin(\omega_0 t) \,\hat{\boldsymbol y} ] ~,
\end{split}
\label{eq:vec_sol_base}
\end{equation}
where the superscript of $\boldsymbol X$ denotes the average particle spin or polarization with respect to the $z$-direction, $\omega_0$ is the oscillating frequency of the dark photon field, and $X(r)$ is a radially symmetric solitonic profile. Given a dark photon of mass $m$, the angular frequency satisfies the conditions $\omega_0<m$ and $m-\omega_0\ll m$. Due to the potential issues caused by vector self-interactions, we assume that dark photon solitons are dominantly supported by gravity and thus neglect self-interactions completely in this work~\cite{Mou:2022hqb, Clough:2022ygm, Coates:2022qia, Capanelli:2024pzd}. In this case, the radial solitonic profile can be approximately described by the fitting formula~\cite{Zhang:2024bjo}
\begin{align}
\label{eq:vec_sol_profile}
X(r) \simeq \bar X \left[1 + 7.89\times 10^{-5}\left(\frac{m M_\mathrm{sol}}{2 M_\mathrm{P}^2}\right)^2 (mr)^2\right]^{-4}\,,
\end{align}
where $M_\mathrm{P} \equiv (8\pi G_\mrm{N})^{-1/2}\approx 2\times 10^{18}\,\mathrm{GeV}$ is the reduced Planck mass, the central field amplitude is
\begin{equation}
    \label{eq:xbar}
    \bar X \simeq 2.98\times 10^{-3} \left(\frac{m M_\mathrm{sol}}{2 M_\mathrm{P}^2}\right)^2 M_\mathrm{P}\,,
\end{equation} 
and $M_\mathrm{sol}$ denotes the soliton mass.

To remain in the non-relativistic regime that allows us to write the base soliton solutions as \cref{eq:vec_sol_base}, the soliton mass must remain below the threshold value~\cite{Salehian:2021khb}
\begin{equation}
M_{\text{sol}} \lesssim 1\times 10^{-4}\, M_\odot\left( \frac{10^{-6}\,\text{eV}}{m} \right)\,.
\label{eq:nonrelbound}
\end{equation}
Moreover, the mass and radius of solitons are related via the expression~\cite{Zhang:2024bjo, Amin:2022pzv, Amin:2023imi}
\begin{equation}
\label{eq:mr_relation}
R_\mathrm{sol} \simeq 197 \frac{ M_\mathrm{P}^2}{m^2 M_\mathrm{sol}} \simeq3\times 10^{-4} R_\odot \left( \frac{10^{-6} \mathrm{eV}}{m} \right)^2 \left( \frac{10^{-9} M_\odot}{M_\mathrm{sol}} \right) ~,
\end{equation}
where $R_\mathrm{sol}$ is defined as the radius that encloses $95\%$ of the total mass, and $R_\odot \approx 6.96\times10^{5}\,\mrm{km}$ is the solar radius. This relation indicates that the heavier the soliton, the smaller its size. 

If the dark photon field couples to electromagnetism via higher-order operators, which we describe in \cref{sec:darkphotonphotoninteraction}, solitons can resonantly produce visible photons through parametric resonance. During this process, the coherent oscillations of the dark photon field generate a periodically varying effective mass that amplifies photon fluctuations, resulting in an exponentially enhanced energy transfer from dark photon modes to electromagnetic ones. A necessary condition for this parametric resonance to occur is that the coupling strength $g$ must be strong enough so that the Bose-Einstein statistics can become effective before the produced photons leave the soliton. Equivalently, the soliton mass must exceed the critical mass value of~\cite{Amin:2023imi}
\begin{equation}
\label{eq:Mc}
M_\mathrm{c} \simeq 42 \frac{M_\mathrm{P}^{4/3}}{g^{2/3} m} 
\simeq 6\times 10^{-10} M_\odot \left( \frac{10^{-6}\,\mathrm{eV}}{m} \right) \left(\frac{10^{-10}\,\mathrm{GeV^{-1}}}{g} \right)^{2/3}\,, 
\end{equation}
where $M_\odot \approx 2 \times 10^{30}\,\mrm{kg}$ is the solar mass.\footnote{The factor of $42$ in this expression is different from that of the same relation in Ref.~\cite{Amin:2023imi}, where this factor was rounded to $100$.} Solitons above this critical mass can trigger explosive bursts of radio-frequency radiation via parametric resonance that are detectable by radio telescopes. 

Therefore, scenarios in which solitons predominantly form with, or evolve toward, masses around this threshold are of significant observational interest. While individual solitons may have masses lying below the parametric resonance condition, major merger events, in which two subcritical solitons with masses around $M_\mrm{c}$ combine, can produce supercritical remnants capable of emitting detectable electromagnetic signals. Such mergers are especially likely within dense astrophysical environments such as galactic bulges, minihalos, and the spiky density profiles surrounding SMBHs. Consequently, accurately determining the soliton merger rate has crucial implications for astrophysical signatures and potential detection strategies. 
A robust prediction of this rate, when combined with the results from the observation of radio telescopes, enables us to impose constraints on the fraction of DM that resides in dark photon solitons $f_\mrm{DM}$, as well as the dark photon-photon coupling strength $g$.

\subsection{Merger Rates in Dark Matter Spikes around Supermassive Black Holes}
\label{sec:merger_rate}

As pioneered in Ref.~\cite{Gondolo:1999ef} and developed by subsequent works \cite{Bertone:2002je, Zhang:2025mdl, Chan:2024yht}, a spiky DM profile is expected to form around SMBHs at the center of galaxies. Due to the significantly enhanced resulting DM density, we expect soliton mergers to occur more frequently around central SMBHs compared with other galactic regions. We calculate the soliton merger rate in these DM spikes below, computing the rate in a typical galaxy without incorporating these spikes in \cref{App:MRMW} for comparison.

Estimating the soliton merger rate is complex; this is partially because not all collisions between solitons lead to mergers. For a merger to take place, the initial total energy needs to be negative so that the system is gravitationally bound. When the spatial separation between the solitons is significantly larger than their individual characteristic radii, one can reasonably approximate the total initial energy of the system as the sum of several distinct components. This includes the total energies associated with each soliton, consisting of their internal energy contributions and their respective kinetic energies. Additionally, we must consider the gravitational interactions that exist between solitons, which can be treated effectively as point masses due to the large distances separating them. Consequently, accounting for these effects, a merger occurs if the relative velocity between solitons $v_\mathrm{rel}$ satisfies~\cite{Hertzberg:2020dbk, Schiappacasse:2025mao}
\begin{align}
\label{eq:mergercond}
v_\mathrm{rel} \lesssim 1.5 ~\mathrm{km\,s^{-1}} \left( \frac{m}{10^{-6}\,\mathrm{eV}} \right) \left( \frac{M_\mathrm{sol}}{10^{-9}\, M_\odot} \right) ~.
\end{align}
Solitons typically have low number densities and carry too much kinetic energy in Milky-Way-like galaxies when ignoring the presence of the spiky DM profiles around SMBHs. This leads to a significant suppression of the galactic merger rate, as has been previously demonstrated for scalar\,\cite{Hertzberg:2020dbk} and tensor\,\cite{Schiappacasse:2025mao} solitons.

However, the presence of DM spikes near SMBHs significantly enhances the soliton merger rate. The expected slope of DM spikes depends on the initial DM profile before the adiabatic formation of the SMBH~\cite{Gondolo:1999ef}. Assuming an initial, pre-formation profile described by the power law $\rho_{\text{DM}} \propto r^{-\gamma}$, the resulting, radial DM spiky profile takes the form~\cite{Zhang:2025mdl}
\begin{equation}
\rho_{\text{DM,sp}}(r)= 0.0263\,M_{\odot}\,\mrm{pc^{-3}} \left[
10^b \left( \frac{M_{\text{SMBH}}}{M_{\odot}} \right)^a 
\times \left( \frac{G_\mrm{N} M_{\text{SMBH}}}{r} \right)^{\omega} \left( 1-\frac{4G_\mrm{N} M_{\text{SMBH}}}{r} \right)^{\eta}\right]
\,,
\label{eq:nspiky}
\end{equation}
where $r$ is the radial distance from the galactic center, $M_\mrm{SMBH}$ is the SMBH mass, and $\{a, b, \omega, \eta\}$ are fit parameters. 

\textcolor{black}{Our analysis adopts a fully relativistic Schwarzschild spike profile, rather than the original semi-relativistic Gondolo--Silk (GS) construction~\cite{Gondolo:1999ef}. The original GS profile provided the canonical benchmark for DM spike formation around a SMBH under the idealized assumption of adiabatic, central black-hole growth, for which the spike develops an inner power-law behavior whose slope is determined by the pre-existing DM distribution. Subsequent work has both extended and revised this picture in several directions, showing that the resulting spike profile is not unique: depending on the black-hole growth history and on the surrounding galactic environment, the inner power-law spike may become either shallower or, in some circumstances, steeper than the canonical GS expectation. In particular, later studies have shown that the spike can be weakened by more realistic SMBH growth histories or sufficiently massive off-center seed formation~\cite{Ullio:2001fb}, by mergers and the associated dynamical heating of the inner halo~\cite{Merritt:2002vj}, and by gravitational scattering off a dense stellar population, which can drive the inner profile toward a shallower cusp or ``mini-cusp''~\cite{Gnedin:2003rj}. Time-dependent evolution in the presence of stars can further deplete the central spike through kinetic heating, self-annihilation, and capture by the SMBH~\cite{Merritt:2003qk,Bertone:2005hw}. More generally, sudden SMBH formation may lead to a shallower spike with slope \(\gamma_{\rm sp}\sim 4/3\)~\cite{Ullio:2001fb}, while stellar heating can weaken the inner spike slope toward \(\sim 3/2\)~\cite{Gnedin:2003rj,Merritt:2003qk,2022PhRvD.106d3018S}. Conversely, mechanisms such as chaotic orbits in triaxial halos~\cite{2004ApJ...606..788M} or gravothermal collapse in self-interacting dark matter~\cite{2000PhRvL..84.5258O} may in some circumstances preserve or enhance the central concentration, yielding a sharper inner profile. While a full treatment of these effects is beyond the scope of this work, their impact on our final results can be partially inferred from our analysis with different slopes $\gamma$.}

\textcolor{black}{Other works generalized the basic spike idea to intermediate-mass black holes and mini-spikes, emphasizing their relevance for indirect-detection searches and, more recently, gravitational-wave probes~\cite{Bertone:2005xz,Eda:2013gg}. On the other hand, for calculations that explicitly involve the strong-field region of a Schwarzschild SMBH, a fully relativistic treatment is more appropriate than the original GS approximation; this motivates our use of the relativistic Schwarzschild spike constructed in the later literature~\cite{Sadeghian:2013laa,Zhang:2025mdl}. Therefore, our choice should be viewed as a representative relativistic benchmark tailored to the present calculation, rather than as a unique or universal astrophysical description of DM spikes around SMBHs.}

\begin{table}[t!]
\centering
\renewcommand{\arraystretch}{1.5}
    \centering
    \begin{tabular*}{\columnwidth}{@{\extracolsep{\fill}}lcccc}
    \toprule\midrule
$\gamma$ & $a$   & $b$  & $\omega$ 
                & $\eta$      \\
    \midrule
    $1$     & $-1.612$ & $31.35$       & $2.09$  & 
    $2.00$   \\
    $1.25$ & $-1.642$ & $32.05$ & $2.11$ & $2.01$  \\            
    $1.5$ & $-1.677$ & $32.81$ & $2.13$ & $2.01$   \\
    $1.75$ & $-1.714$ & $33.70$ & $2.16$ & $2.04$  \\
    \midrule\bottomrule
  \end{tabular*}
  \caption{Best-fit parameters for \cref{eq:nspiky} for the spiky dark matter profile $\rho_\mrm{DM, sp}$ formed around supermassive black holes of masses between $10^4 M_{\odot}$ and $10^9 M_{\odot}$. Results are from Ref.~\cite{Zhang:2025mdl}.}
    \label{tab:best-fit-params}
  \end{table}

\begin{figure}
    \centering
  \includegraphics[width=0.93\textwidth]{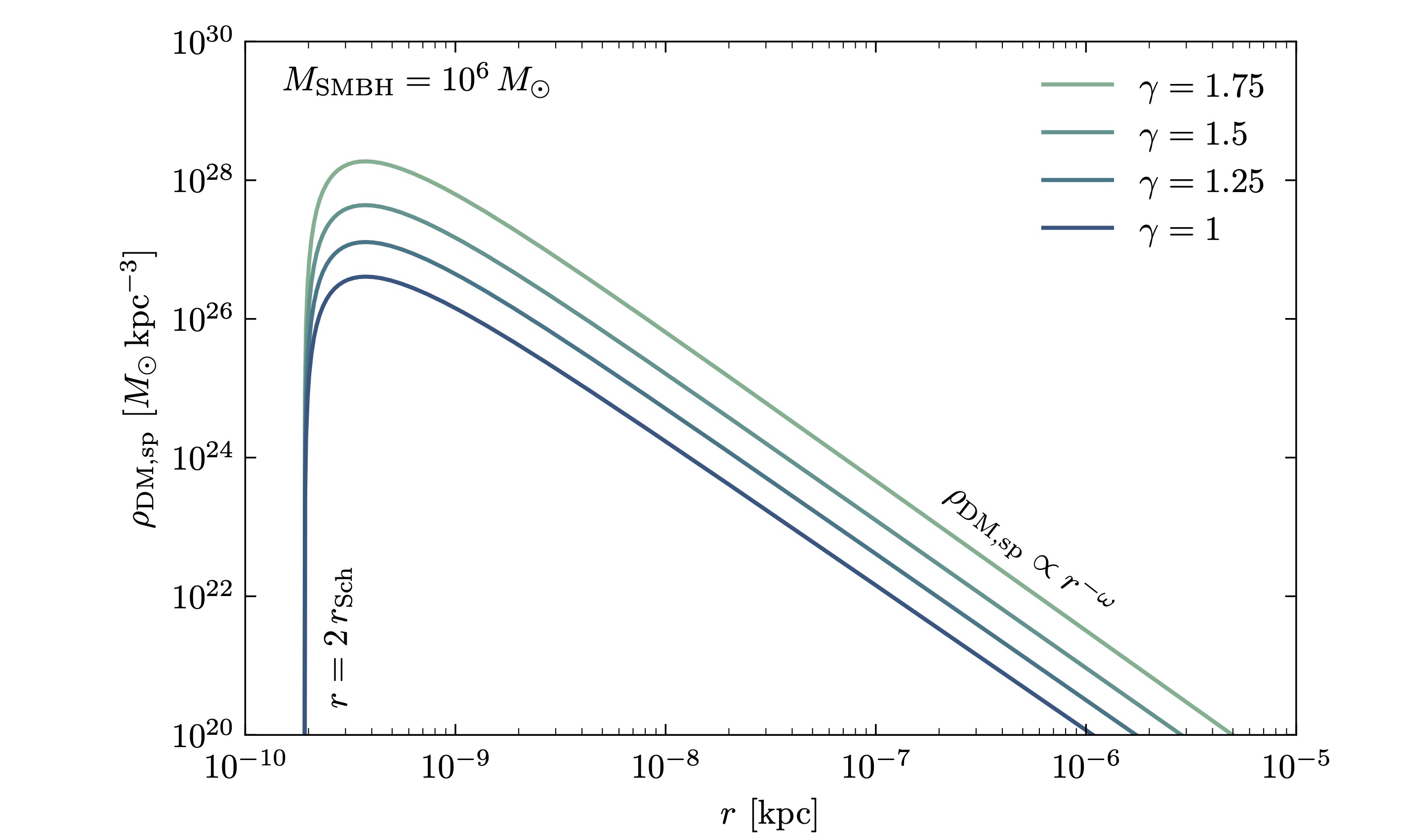}
    \caption{\justifying The spiky dark matter density profile $\rho_\mrm{DM,sp}(r)$ with radial distance $r$ around a supermassive black hole of mass $M_\mathrm{SMBH} =10^6\,M_\odot$. Each profile is computed for different initial mass dark halo functions, governed by the index parameter $\gamma$. We show the results for $\gamma \in \{1,1.25,1.5,1.75\}$. The density function commences at twice the Schwarzschild radius $r = 2r_\mrm{Sch}$ and its tail falls off as $\rho_\mrm{DM,sp} \propto r^{-\omega}$, with $\omega$ given in \cref{tab:best-fit-params} for each value of $\gamma$.}
    \label{fig:rhosp}
\end{figure}

\begin{figure}
    \centering
    \begin{subfigure}{0.495\textwidth}   \includegraphics[width=0.93\textwidth]{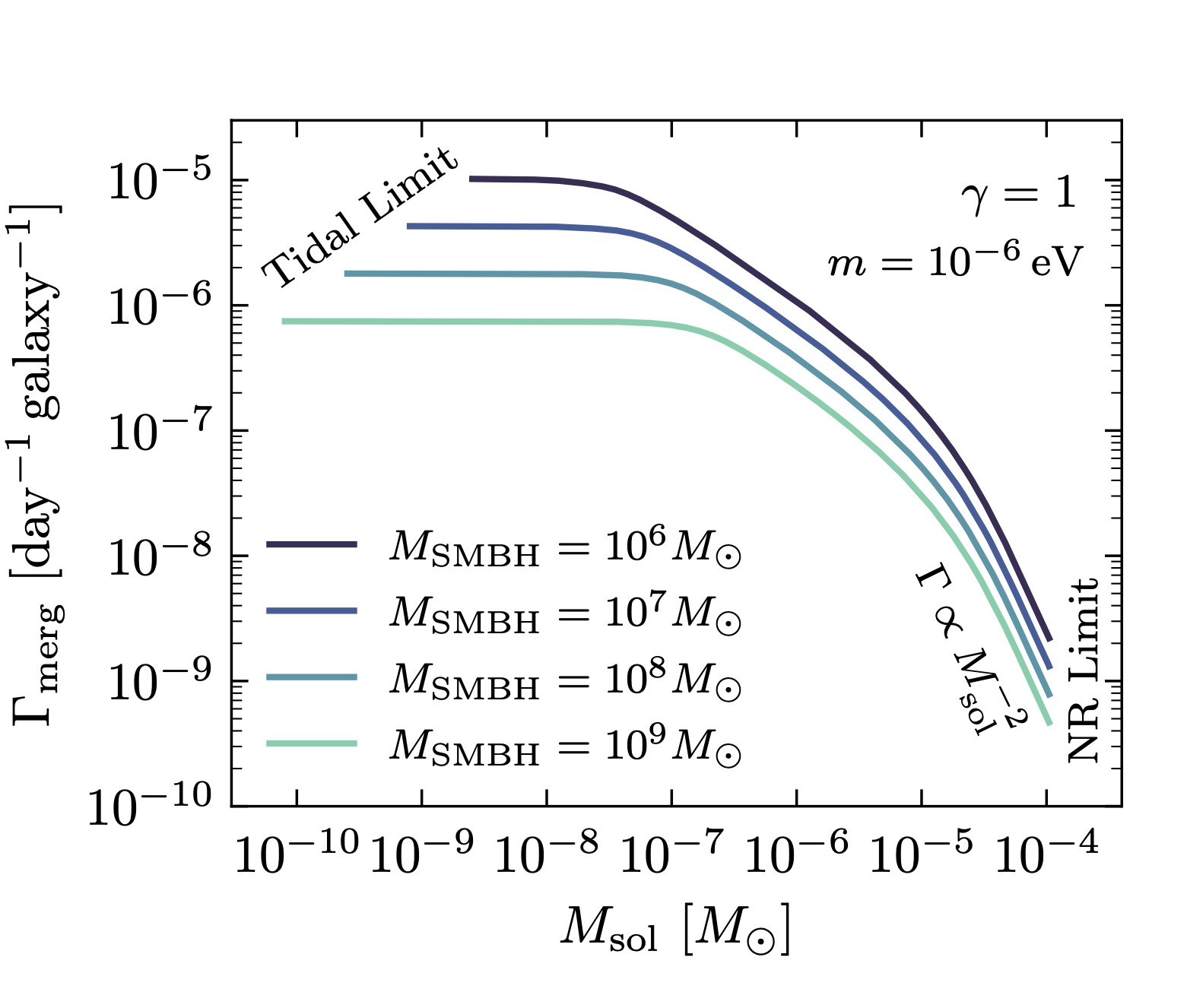}
    \end{subfigure}
    \begin{subfigure}{0.495\textwidth}
    \includegraphics[width=0.93\textwidth]{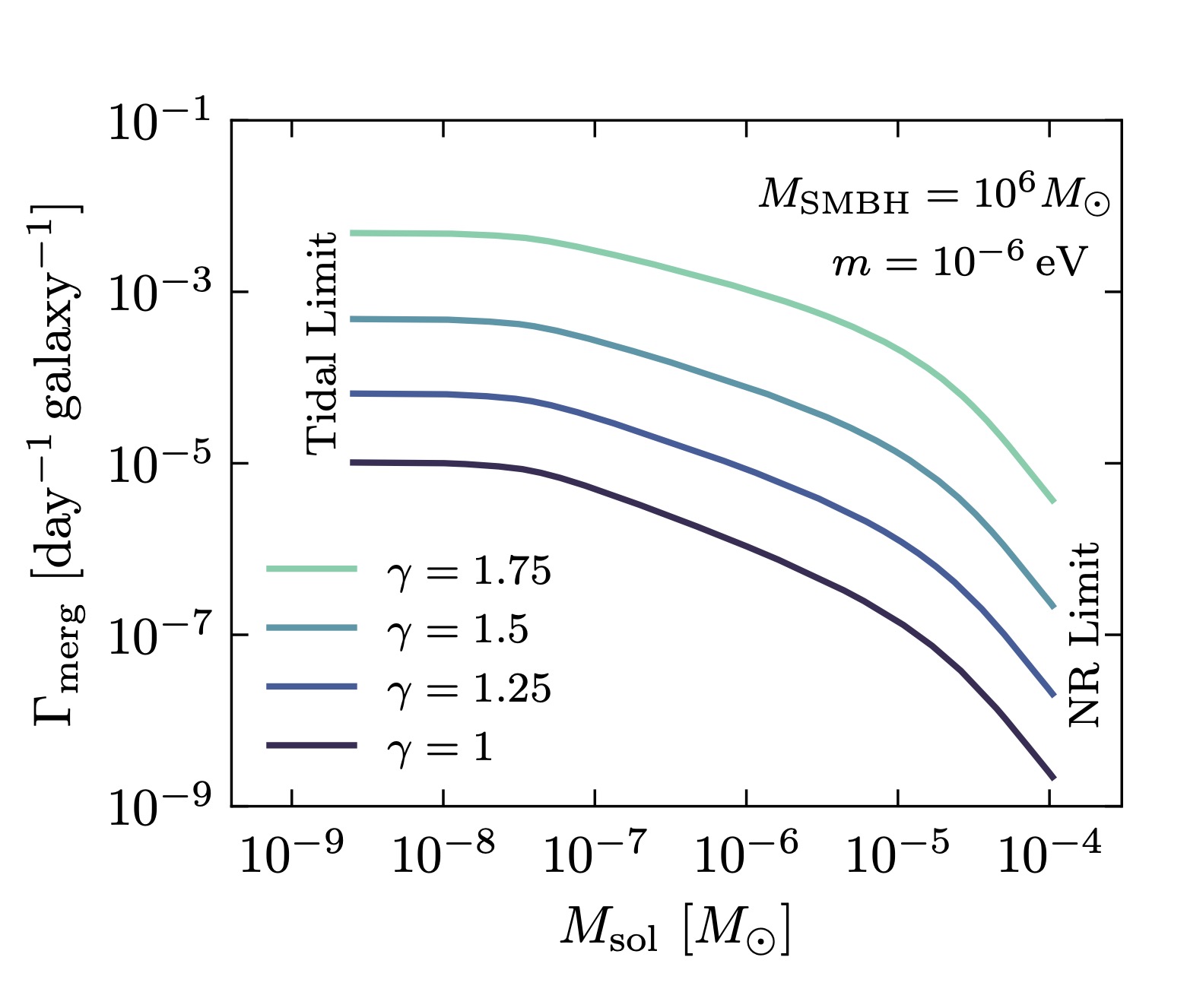}
    \end{subfigure}
    \caption{\justifying The soliton merger rate $\Gamma_\mrm{merg}$ with soliton mass $M_\mrm{sol}$ in the presence of the spiky dark matter profiles formed around supermassive black holes at galactic centers. Here we assume that vector solitons compose all of the dark matter, such that $f_\mrm{DM} = 1$, and show the result for the dark photon mass $m=10^{-6}\,\mrm{eV}$. The accessible soliton masses are limited from below by the constraint that they must be tidally stable around the supermassive black hole (\cref{eq:msol_tidal}) and from above by the condition that we must remain in the non-relativistic (NR) regime (\cref{eq:nonrelbound}) \textbf{Left:} The merger rate assuming a power-law index of $\gamma = 1$ for the initial dark matter density profile and for varying supermassive black hole mass. \textbf{Right:} The merger rate assuming a supermassive black hole of mass $M_\mrm{SMBH} = 10^{6}M_\odot$ for varying values of the power-law index $\gamma$.}
    \label{fig:mrfixedSMBH}
\end{figure}

The relation \eqref{eq:nspiky} is valid for $2r_{\text{Sch}} \lesssim r \lesssim r_{\text{sp}}$, where $r_{\text{Sch}} \equiv 2G_\mrm{N} M_{\text{SMBH}}$ is the Schwarschild radius, $2 r_{\text{Sch}}$ is the marginally bound orbit\,\cite{Bardeen:1972fi}, and $r_{\text{sp}}$ is the spiky radius, which depends on  $\gamma$. The best-fit values for $\{a,b,\omega,\eta\}$ for SMBHs with masses ranging from $10^4\, M_\odot$ to $10^{9}\,M_\odot$ for initial power-law indices $\gamma \in \{1,1.25,1.5,1.75\}$ are shown in \cref{tab:best-fit-params}\,.\footnote{$N$-body DM simulations indicate that the value of $\gamma$ falls within the range $0.9 < \gamma < 1.5$~\cite{Diemand:2008in, 2010MNRAS.402...21N}. However, the collapse of baryonic matter into a baryonic disk can lead to a steeper power-law index. For example, in the innermost regions of disk galaxies, $N$-body/hydrodynamical simulations including DM, stars, and gas have found a power-law index within the range of $1.7 < \gamma < 2.1$~\cite{Gustafsson:2006gr}.} For the case of Milky-Way-like galaxies, we have $r_{\text{h}} \lesssim r_{\text{sp}} \lesssim\, \mathcal{O}(10)\, r_{\text{h}}$ for $1 \leq \gamma \leq 1.75$\,\cite{Gondolo:1999ef}, where $r_{\text{h}}$ is the radius of gravitational influence of the black hole $r_\mrm{h}  \equiv G_\mrm{N} M_{\text{SMBH}}/\sigma_{*}^2$\,\cite{1972ApJ...178..371P}, with $\sigma_{*}$ the stellar velocity dispersion of the host bulge. We take a conservative approach and calculate the merger rates within the inner spherical galactic region,  $2r_{\text{Sch}} \lesssim r \lesssim r_{\text{h}}$.
We show the DM densities for spiky profiles with varying galactic radius around an SMBH of mass $10^6\,M_\odot$ for choices of the power-law index $\gamma \in \{1,1.25,1.5,1.75\}$ in \cref{fig:rhosp}. Larger values of $M_{\text{SMBH}}$ and $\gamma$ lead to more concentrated density profiles across all $r$.\footnote{ \textcolor{black}{
The spiky profile is not fixed by $M_{\rm SMBH}$ alone: it depends on the pre-formation inner halo profile. In our modeling, this uncertainty is parameterized by the initial inner slope $\rho_{\rm DM}\propto r^{-\gamma}$, which controls the resulting spike through Eq.\,\eqref{eq:nspiky} (see also Table\,\ref{tab:best-fit-params}). We therefore treat $\gamma$ as an astrophysical systematic, though, unless otherwise stated, we adopt $\gamma=1$ as a conservative benchmark. Larger values of $\gamma$ produce denser spikes and hence larger merger rates/stronger constraints (see, e.g., Fig.\,\ref{fig:rhosp} and the right panel of Fig.\,\ref{fig:mrfixedSMBH}). For completeness, we show results for other $\gamma$ values in \cref{sec:app-all_gamma}.}}

While the increased density profiles around SMBHs lead to enhanced soliton merger rates, solitons around these extreme astrophysical environments risk being tidally disrupted and torn apart by the gravitational influence of the black hole. Tidal disruption occurs if they remain within the Roche radius of the SMBH, given by\footnote{\textcolor{black}{The soliton mass–radius relation in Eq.\,\eqref{eq:mr_relation} can also be derived in a Hamiltonian framework, in which the non-relativistic Hamiltonian would split naturally into kinetic and self-gravitational contributions\,\cite{Schiappacasse:2017ham}. Vector solitons are long-lived configurations sustained by a balance between inward self-gravity and outward kinetic (gradient) pressure. Consequently, vector solitons—like other self-gravitating solitonic structures such as scalar and tensor solitons—can be treated as compact astrophysical objects when assessing the impact of external tidal fields from nearby bodies (e.g. stars), as has been done in the literature\,\cite{Glennon:2022huu, Edwards:2018ccc, Raby:2016deh, Schiappacasse:2025mao, Enrico:2025}.}}
\begin{equation}
r_\mathrm{tidal}  \simeq R_\mathrm{sol} \left(\frac{M_\mathrm{SMBH}}{M_\mathrm{sol}} \right)^{1/3} 
\simeq 30\,R_\odot \left( \frac{M_\mathrm{SMBH}}{10^6\, M_\odot} \right)^{1/3}
\left( \frac{10^{-9}\,M_\odot}{M_\mathrm{sol}} \right)^{4/3} \left( \frac{10^{-6}\,\mathrm{eV}}{m} \right)^2\,, 
\end{equation}
where we have used the soliton mass-radius relation given in \cref{eq:mr_relation}. Solitons in the spiky profile can remain stable against the tidal disruption from a central SMBH if the Roche radius is less than the inner boundary of the profile, given by twice the Schwarzschild radius: $r_\mrm{tidal} < 2 r_\mrm{Sch} =4 G_\mrm{N} M_\mathrm{SMBH}$. To defend against tidal disruption, solitons must therefore have masses,\footnote{Other disruptive events such as soliton-star encounters are expected to be subdominant compared to soliton mergers due to the higher event rate of the latter. Detailed analyses that include such effects can be found in Ref.~\cite{Tinyakov:2015cgg, 2017JETP..125..434D, Blinov:2019jqc, Kavanagh:2020gcy,Hertzberg:2019exb,Nurmi:2021xds}.}
\begin{equation}
    \label{eq:msol_tidal}  
    M_\mrm{sol} \gtrsim 3 \times 10^{-9}\,M_\odot\,\left(\frac{10^6 \, M_\odot}{M_\mrm{SMBH}}\right)^{1/2} 
     \left(\frac{10^{-6}\,\mrm{eV}}{m}\right)^{3/2}\,.  
\end{equation}

For such solitons, the total merger rate at galactic centers for a fixed SMBH mass can be estimated as\,,\footnote{\textcolor{black}{In our analysis, we adopt a phenomenological description in which solitons constitute a fixed fraction \(f_{\rm DM}\) of the dark matter mass density locally, so that the soliton number density is
\(n_{\rm sol}(r)= f_{\rm DM}\,\rho_{\rm DM,sp}(r)/M_{\rm sol}\).
This assumption means that, \emph{to leading order}, solitons behave as a collisionless macroscopic
DM subcomponent that is redistributed by the same gravitational potential that shapes the spike.
It is also consistent with the physical picture in which solitons are typically nucleated before SMBH growth and then respond adiabatically to the gradually deepening central potential (subject to tidal survival, which we already impose via the Roche criterion and the tidal limit shown in Fig.\,\ref{fig:mrfixedSMBH}). This is, for example, the case for vector solitons produced by the collapse of Hubble-scale inhomogeneities around radiation–matter equality \cite{Gorghetto:2022sue}.}} 
\begin{equation}
\Gamma_{\text{merg}} = \int^{r_{\text{h}}}_{2r_{\text{Sch}}}\frac{4\pi r^2}{2} \langle \sigma_{\text{eff}}\,v_{\text{rel}}  \rangle_{\text{merg}}  
\left( \frac{\rho_{\mrm{DM, sp}}(r)f_{\text{DM}}}{M_{\text{sol}}} \right)^2\,\mrm{d}r\,, 
\label{eq:merger}
\end{equation}
where the gravitational influence radius for each SMBH mass is calculated using the mass-velocity-dispersion relation $(M_{\text{SMBH}},\sigma^* )$ given in~\cite{2000ApJ...539L...9F}, such that $\log_{10}(M_{\text{SMBH}}/M_{\odot})=8.12 - 4.24 \log_{10}(200\, \mrm{km\,s^{-1}}/\sigma^*)$. The effective merger cross section averaged over the relative velocities is calculated as in \cref{eq:averagevrel}, which we quote here for clarity:
\begin{equation}
\langle \sigma_{\text{eff}}\,v_{\text{rel}}  \rangle_{\text{merg}} = 
\int_0^{\text{min}\left(2v_{\text{esc}},v^*_{\text{rel}}\right)}   \funop{p({v_{\text{rel}}})}\sigma_{\text{eff}}\,v_{\text{rel}}\dd v_{\text{rel}}\,,
\label{eq:averagevrelmaintext}
\end{equation}
where $v^*_{\text{rel}}$ saturates the inequality given in \cref{eq:mergercond}, the escape velocity is approximated to be $v_{\text{esc}} \simeq (2G_\mrm{N} M_{\text{SMBH}}/r)^{1/2}$, and the effective cross section $\sigma_{\text{eff}}\,v_{\text{rel}}$ is given by the geometrical cross section times the gravitational focusing enhancement as in \cref{sigmaeff}. The upper limit of integration  imposes the merger condition given in \cref{eq:mergercond}, ensuring that soliton collisions lead to mergers.
The probability density function governing the relative velocities of solitons $\funop{p({v_{\text{rel}}})}$ is taken to be a 3D isotropic Maxwell-Boltzmann distribution, as in \cref{eq:Pvrel}. However, the associated 3D isotropic relative velocity dispersion $\sigma_{\text{rel}}$ is obtained by generalizing \cref{eq:Jeq}.

To find $\sigma_\mathrm{rel}$, we first take \cref{eq:nspiky} to hold a power-law behavior of the form  $\rho_{\text{DM,sp}} \propto r^{-\omega}$, where the values of $\omega \equiv \omega(\gamma)$ are listed in \cref{tab:best-fit-params} and computed in Ref.~\cite{Zhang:2025mdl}. We write the spherical Jeans equation within the spiky region assuming that the soliton dispersion velocity is isotropic, such that $\sigma_r = \sigma_\theta=\sigma_\phi \equiv \sigma_{\text{DM}}$, where $\sigma_\mrm{DM}$ is the DM velocity dispersion. We then have that
\begin{equation}
\frac{\partial(r^{-\omega}\sigma_{\text{DM}}^2)}{\partial r} + r^{-\omega}\frac{\partial \Phi_\mrm{N}(r)}{\partial r} =0 \,,
\end{equation}
where $\Phi_\mrm{N}(r) \equiv -G_\mrm{N} M_{\text{SMBH}}/r$ is the gravitational potential at radius $r$. The solution of this equation is
\begin{equation}
\sigma_{\text{DM}}(r) = \left(1+\omega\right)^{-1/2}\left( \frac{G_\mrm{N} M_{\text{SMBH}}}{r}\right)^{1/2}\,,
\label{eq:Jeq2}
\end{equation}
and we use $\sigma_\mathrm{rel} = \sqrt{2} \sigma_\mathrm{DM}$.
\textcolor{black}{Note that the relative velocity distribution used to compute \(\langle \sigma_{\rm eff}v_{\rm rel}\rangle_{\rm merg}\), Eq.~(\ref{eq:averagevrelmaintext}), considers enhanced velocities in the inner region and their dependence on the spike slope.
Any additional departures from a Maxwell--Boltzmann shape (e.g.~ anisotropy or non-thermal tails) would be a higher-order
effect on top of a dispersion that is already fixed self-consistently by the spiky profile and the SMBH potential.}
%
%
%
 
We show the soliton merger rate around SMBHs in \cref{fig:mrfixedSMBH}. We show this as a function of the soliton mass with both varying SMBH mass for a fixed initial power-law index of $\gamma = 1$ and varying power-law index $\gamma$ for a fixed SMBH mass of $M_\mrm{SMBH} = 10^{6}M_\odot$. In either case, we fix the dark photon mass to be $m = 10^{-6}\,\text{eV}$ and normalize the DM fraction in vector solitons to be unity, such that $f_\text{DM} = 1$. Each curve commences at the tidally limited soliton mass, given by \cref{eq:msol_tidal}, and terminates at the soliton mass upper bound from the validity of the non-relativistic approximation, given by \cref{eq:nonrelbound}.

The larger the SMBH mass, the larger the gravitational influence radius and the DM spiky profile at fixed radius, shown in \cref{eq:nspiky}. However, larger SMBH masses also lead to higher soliton dispersion relative velocities, given in \cref{eq:Jeq2}. While the first two factors tend to enhance the merger rate, the third factor tends to suppress it. The rise in $\sigma_{\text{DM}}$ leads to an exponential suppression of the 3D isotropic Maxwell-Boltzmann distribution function for the soliton relative velocities. This suppression effect outweighs the increase in the gravitational influence radius and DM density. Therefore, the ultimate result is that the merger rate decreases
as $M_{\text{SMBH}}$ increases for fixed $M_{\text{sol}}$. Moreover, since $\Gamma_{\text{merg}} \propto \rho^2_{\text{DM,sp}}$, the merger rate increases as the spiky DM density profile increases for fixed SMBH, soliton, and dark photon masses. The profile increases as $\gamma$ grows, and this behavior is seen in the right panel of \cref{fig:mrfixedSMBH}.

As explained in \cref{App:MRMW} and visualized in \cref{fig:MR2}, \cref{eq:merger} possesses two distinguishable regimes. When the upper limit of integration in \cref{eq:averagevrelmaintext} is dominated by the escape velocity of solitons, the effective cross section is controlled by the geometrical cross section, such that $\Gamma_{\text{merg}} \propto (\sigma_{\text{eff}}\,v_{\text{rel}})M^{-2}_{\text{sol}} \propto \mrm{constant}$. This explains the flat behavior of the curves in \cref{fig:mrfixedSMBH}. Conversely, when the upper limit of integration is instead dominated by $v^*_{\text{rel}}$, it is the gravitational focusing that governs the effective cross section, resulting in the behavior 
$\Gamma_{\text{merg}} \propto (\sigma_{\text{eff}}\,v_{\text{rel}})M^{-2}_{\text{sol}} \propto M^{-2}_{\text{sol}}m^{-2}$, as is indicated in \cref{fig:mrfixedSMBH} (left panel).

The effect of the spiky DM profile can be seen by comparing our result to our calculation for the Milky Way given in \cref{App:MRMW}. In the latter computation, the results of which are given in \cref{fig:MR1}, we have derived the soliton merger rate for the entire Milky Way DM halo. As shown in \cref{fig:MR2}, the transition between the regimes where the escape velocity or \(v^*_{\text{rel}}\) dominates (across most of the galactic radius range) spans only one order of magnitude in the quantity \((M_{\text{sol}}m)\). This characteristic causes the flat behavior of the merger rate curve to quickly exhibit the dependence \(\Gamma_{\text{merg}}{\text{sol}} \propto M^{-2}_{\text{sol}}m^{-2}\) as soon as \(M_{\text{sol}}m \gtrsim \mathcal{O}(1) \times 10^{-13}\,M_{\odot}\,\text{eV}\). In contrast, the transition between both regimes in \cref{fig:mrfixedSMBH} (left panel) spans several orders of magnitude in the quantity \((M_{\text{sol}}m)\) since the merger rate is calculated only within the gravitational influence radius.\footnote{This cut-off for the radius is justified by the fact that the galactic merger rate is dominated by the DM spike around the SMBH.} As a result, the transition between the regimes where escape velocity or \(v^*_{\text{rel}}\) dominates is less pronounced as the quantity \((M_{\text{sol}}m)\) varies.

To estimate the total (local and nonlocal) dark photon soliton merger rate in spiky DM centers around SMBHs at certain redshifts $z^*$, we utilize the SMBH mass function $\Phi(M_{\text{SMBH}},z)$, defined as the number of SMBHs per comoving volume per logarithmic mass interval, for $0 \leq z \leq z^*$. 
%
\begin{figure}[t]
    \centering
    \includegraphics[width=0.93\textwidth]{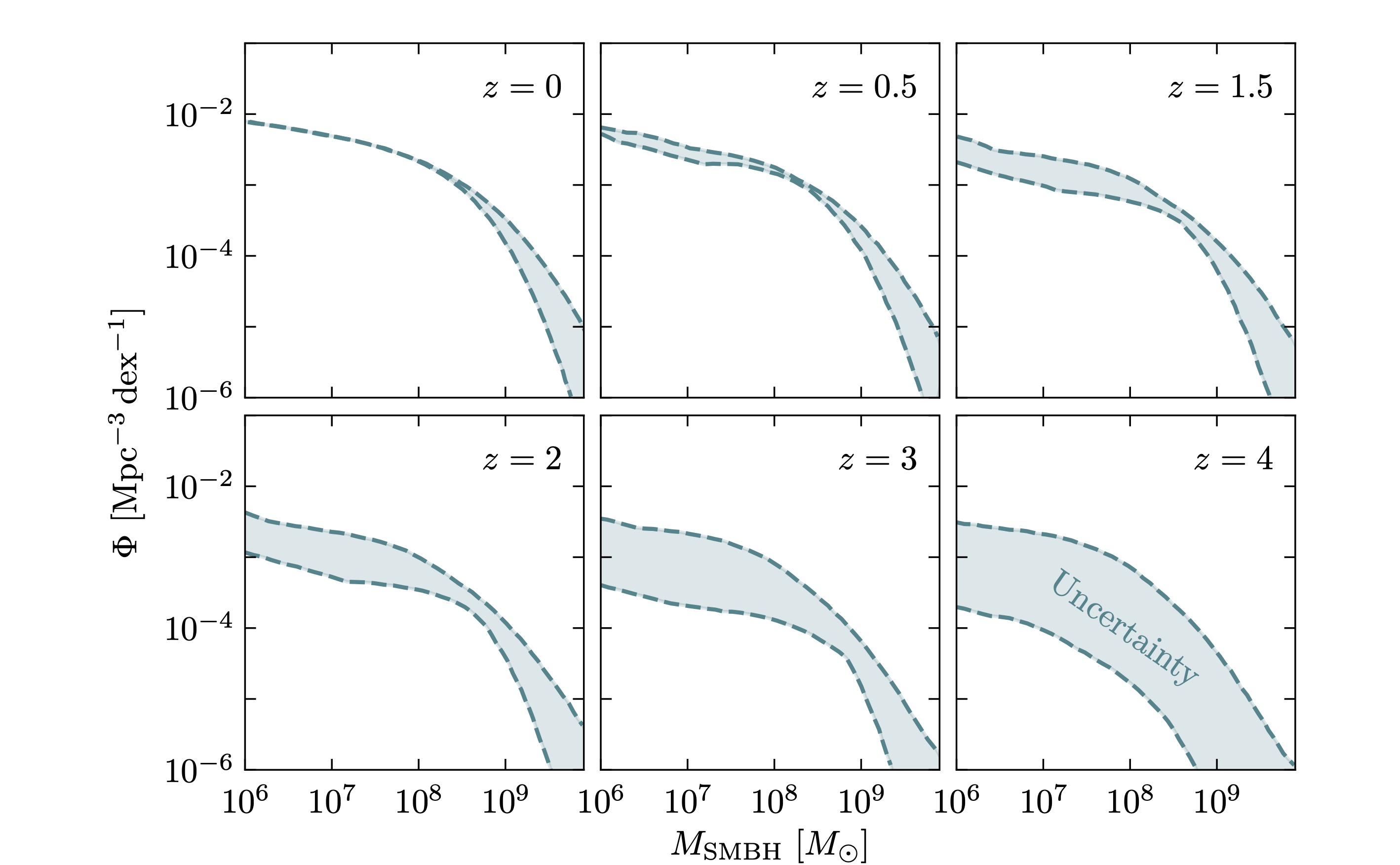}
    \caption{\justifying The supermassive black hole mass function $\Phi$ with black hole mass $M_\mrm{SMBH}$ at varying redshifts $z$. These are computed in Ref.~\cite{Tucci:2016tyc} for different model assumptions, leading to an uncertainty in the ultimate result. This is illustrated via the shaded bands.}
    \label{fig:PhiSMBH}
\end{figure}
%
Generally, while it is possible to measure the SMBH mass of several galaxies locally through different techniques\,\cite{2013ARA&A..51..511K}, this is impossible nonlocally, and the observed population only refers to the active ones. The mass function of local SMBHs has been estimated multiple times using various methods that consider the observed relationships between SMBH masses and the properties of their host galaxies, such as bulge mass, luminosity, and stellar velocity dispersion\,\cite{2009ApJ...690...20S, 2009MNRAS.400.1451V, 2011ApJ...742...33L, 2013MNRAS.428..421S, 2014ApJ...786..104U}. All results are consistent with each other despite uncertainties regarding the actual local SMBH function. 

In Ref.\,\cite{Tucci:2016tyc}, the convolution of a Schechter function with a Gaussian scatter of $0.3\,\mrm{dex}$ and $0.5\,\mrm{dex}$ is used as an analytical expression, aiming to capture the current uncertainties in the local SMBH function for higher masses. Here, the unit of `dex' represents \textit{decimal exponent}, such that a change of $1\,\mrm{dex}$ corresponds to a change by a factor of $10$ in mass. Additionally, they estimate the evolution of SMBHs over time, tracing their development from the present back to a redshift of $z \leq 4$ using the continuity equation. They observe a dependence between the obtained mass function and the SMBH average radiative efficiency, the local SMBH mass function, and the SMBH Edddington ratio. We show their results on the upper and lower limits regarding the predicted SMBH mass function at certain redshifts in \cref{fig:PhiSMBH}. We see that the SMBH mass function evolves only until $M_{\text{SMBH}} \lesssim 10^8 M_{\odot}$ for $z \lesssim 1$, but heavier SMBHs undergo a strong evolution at higher redshifts. 

Using these results, we estimate the merger rate of dark photon solitons around SMBHs at the center of galaxies.
We compute the total merger rate per unit volume and per unit time for the redshift and mass ranges $0 \leq z \leq 4$ and $M^{\text{min}}_{\text{SMBH}}  \leq M_{\text{SMBH}} \leq M^{\text{max}}_{\text{SMBH}}$, respectively, with
$M^{\text{min}}_{\text{SMBH}}  =10^5\,M_{\odot}$ and $M^{\text{max}}_{\text{SMBH}}  =8 \times 10^9\,M_{\odot}$. We have
\begin{equation}
\Gamma^{\text{TOTAL}}_{\text{merg}} = \int_0^4 \int_{\text{log}(M^{\text{min}}_{\text{SMBH}}/M_{\odot})}^{\text{log}(M^{\text{max}}_{\text{SMBH}}/M_{\odot})} \Gamma_{\text{merg}} (M_{\text{SMBH}},\gamma) 
\Phi(M_{\text{SMBH}},z)\dd\text{log}_{10}M_{\text{SMBH}}\dd z\,,
\label{eq:MRtotal}
\end{equation}
where $\Gamma_{\text{merg}} (M_{\text{SMBH}},1)$, for example, is shown in \cref{fig:mrfixedSMBH}.
The mass integration is in the logarithmic scale since the mass function $\Phi$ is defined with respect to logarithmic mass intervals.\footnote{Note that we have extended the validity of Eq.\,(\ref{eq:nspiky}) to  $M_{\text{SMBH}} = O(10^{10})\,M_{\odot}$. We have performed a preliminary numerical estimate that indicates that Eq.\ (\ref{eq:nspiky}) holds well for this range of SMBH masses.}
%
\begin{figure}
    \centering
    \includegraphics[width=0.93\textwidth]{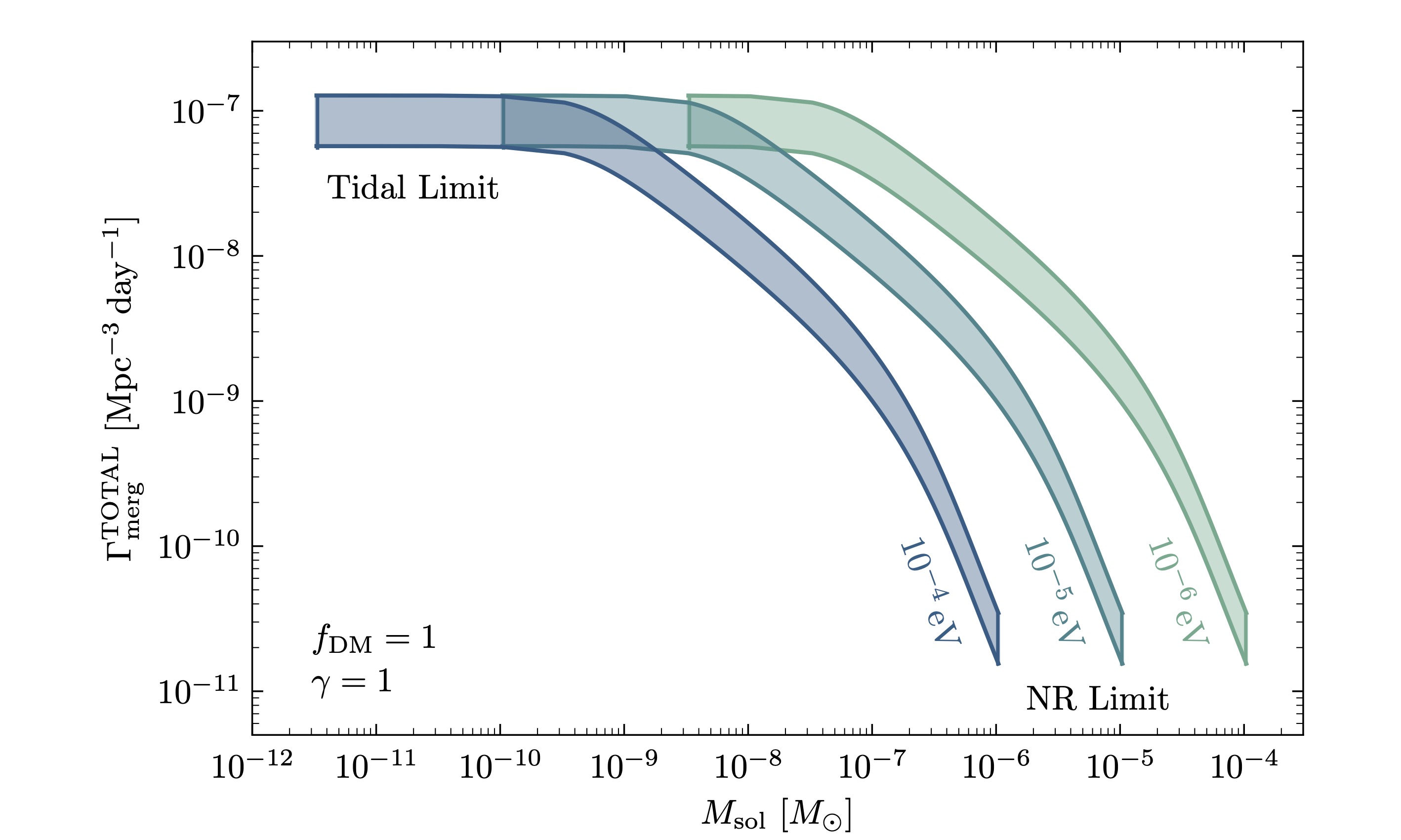}
    \caption{\justifying The total merger rate of vector solitons in dark matter spikes per unit volume $\Gamma_\mrm{merg}^\mrm{TOTAL}$ integrated over supermassive black holes with masses in the range $10^6\, M_{\odot} \leq M_{\text{SMBH}}\leq 8 \times 10^9\, M_{\odot}$  from $z=0$ to $z=4$ in terms of the soliton mass $M_\mrm{sol}$. Each colored band assumes a particular value for the dark photon mass, which we show for the range $m \in \{10^{-6}\,\mrm{eV}, 10^{-5}\,\mrm{eV}, 10^{-4}\,\mrm{eV}\}$. Here we assume that dark photon solitons compose all of the dark matter, such that $f_\mrm{DM}=1$, and we take $\gamma=1$ for the initial dark matter profile, including the uncertainties in the supermassive black hole mass functions shown in \cref{fig:PhiSMBH}. The lower bound on the soliton mass is governed by the limit in which they can be tidally disrupted by supermassive black holes (\cref{eq:msol_tidal}), and the upper bound is dictated by remaining in the non-relativistic (NR) limit (\cref{eq:nonrelbound}). We show the results for other $\gamma$ values in \cref{sec:app-all_gamma}.}
    \label{fig:MR}
\end{figure}
%

We show the total merger rate of vector solitons in terms of the soliton mass within DM spikes around SMBH with masses in the range $10^6\, M_{\odot} \leq M_{\text{SMBH}}\leq 8 \times 10^9\, M_{\odot}$  from $z=0$ to $z=4$ in \cref{fig:MR}. Each band corresponds to a particular value for the mass of dark photons. For small soliton masses, we have $\Gamma_{\text{merg}}^{\text{TOTAL}} \lesssim 10^{-7}f^2_{\text{DM}}\,\text{Mpc}^{-3}\,\text{day}^{-1}$.

\section{Radiation from Parametric Resonance}
\label{sec:param-res}

The coherent, periodically oscillating nature of the dark photon vector field can lead to a phenomenon known as parametric resonance in dark photon solitons. This occurs if dark photons couple to electromagnetism and if the soliton mass exceeds the critical threshold given in \cref{eq:Mc}. The parametric resonance of vector solitons can then lead to bursts of high-energy radio frequency radiation detectable by radio telescopes. 

\subsection{Dark Photon-Photon Coupling}
\label{sec:darkphotonphotoninteraction}
Consider an action $\mathcal{S}$ that includes the dark sector, the electromagnetic sector, and their interactions. We can write this action as
\begin{equation}
\label{eq:action}
\mathcal{S} \equiv  \int \dd^4 x \sqrt{-g} \biggl[ 
\frac{1}{2} M_{\text{P}}^2 R - \frac{1}{4} X_{\mu\nu} X^{\mu\nu} - \frac{1}{2} m^2 X_\mu X^\mu 
 - \frac{1}{4} F_{\mu\nu} F^{\mu\nu} + \mathcal{L}_\mathrm{int} \biggr]\,,
\end{equation}
where $R$ is the Ricci scalar, and where we respectively denote the photon and dark photon fields as $A_{\mu}$ and $X_{\mu}$. Their corresponding field strength tensors are given by $F_{\mu\nu} \equiv \partial_\mu A_\nu - \partial_\nu A_\mu$ and $X_{\mu\nu} \equiv \partial_\mu X_\nu - \partial_\nu X_\mu$. The Lagrangian term $\mathcal{L}_\mathrm{int}$ encodes the interactions between the two sectors. In the framework of effective field theory (EFT), we may write $\mathcal{L}_\mathrm{int} \equiv g^2 \mathcal{O}_i$, where $\mathcal{O}_i$ is an effective interaction operator and $g$ is the effective gauge coupling strength. The leading-order gauge-invariant operators that give rise to parametric resonance of photons include~\cite{Amin:2023imi}
\begin{equation}
\begin{split}
\mathcal{O}_1 &\equiv -\frac{1}{2} F_{\mu\nu} \tilde{F}^{\mu\nu} X_{\alpha} X^{\alpha} 
\simeq  \ 
    2 (\Evec \cdot \Bvec) (\Xvec \cdot \Xvec)\,,\\ 
\mathcal{O}_2 &\equiv-\frac{1}{2} F_{\mu\nu} F^{\mu\nu} X_{\alpha} X^{\alpha}
\simeq  \ 
    (\Evec \cdot \Evec) (\Xvec \cdot \Xvec) - (\Bvec \cdot \Bvec) (\Xvec \cdot \Xvec)\,,\\  
\mathcal{O}_3 &\equiv F_{\mu\rho} F^{\nu\rho} X^\mu X_\nu 
\simeq   
    (\Bvec \cdot \Bvec) (\Xvec \cdot \Xvec) - (\Evec \cdot \Xvec)^2 - (\Bvec \cdot \Xvec)^2\,,\\
\mathcal{O}_4 &\equiv \tilde{F}_{\mu\rho} \tilde{F}^{\nu\rho} X^\mu X_\nu 
  \simeq  \ 
    (\Evec \cdot \Evec) (\Xvec \cdot \Xvec) - (\Evec \cdot \Xvec)^2 - (\Bvec \cdot \Xvec)^2\,,
\end{split}
\label{eq:Operators}
\end{equation}
where we have neglected terms dependent on $X_{0}$ and $\nabla X_\mu$ since they are suppressed by the low, non-relativistic velocities of cold dark photons, $v \sim 10^{-3}$. We have also included how each operator can be approximately written in terms of the electric field $\bvec{E}$, magnetic field $\bvec{B}$, and dark photon $3$-vector field $\bvec{X}$.

The validity of the EFT requires that the coupling constant and the dark photon field amplitude remain small, such that $g^2 \bar{X}^2 \ll 1$, with $\bar{X}$ given in \cref{eq:xbar}~\cite{Amin:2023imi}. This justifies us neglecting higher-order operators related to parametric resonance. This condition also ensures that the contributions from dimension-$6$ operators to lower-dimensional operators, induced by the nonzero vacuum expectation value of the dark photon field, remain negligible.

Our EFT description is independent of any chosen ultraviolet (UV) embedding. However, we note that the operators listed in \cref{eq:Operators} can originate from a simple, renormalizable theory at the UV scale. For instance, the operator $\mathcal{O}_2$ can be generated at a mass scale $M$ where new physics is expected to emerge via the dimension-$8$ operator $\mathcal{L}_{\mrm{int}} = -\frac{1}{8}M^{-4}|D_{\alpha}\phi|^2F_{\mu\nu}F^{\mu\nu}$. Here, $D_{\alpha}\phi$ represents the covariant derivative of a dark Higgs field $\phi$, while $X_{\mu}$ is a novel field associated with a dark $U(1)$ gauge symmetry. The dimension-$6$ operator would then be generated by the non-zero vacuum expectation value of the dark Higgs field~\cite{Amin:2023imi}. 

\begin{figure*}[t]
    \centering
    \begin{subfigure}[t]{0.495\textwidth}
    \includegraphics[width=0.93\textwidth]{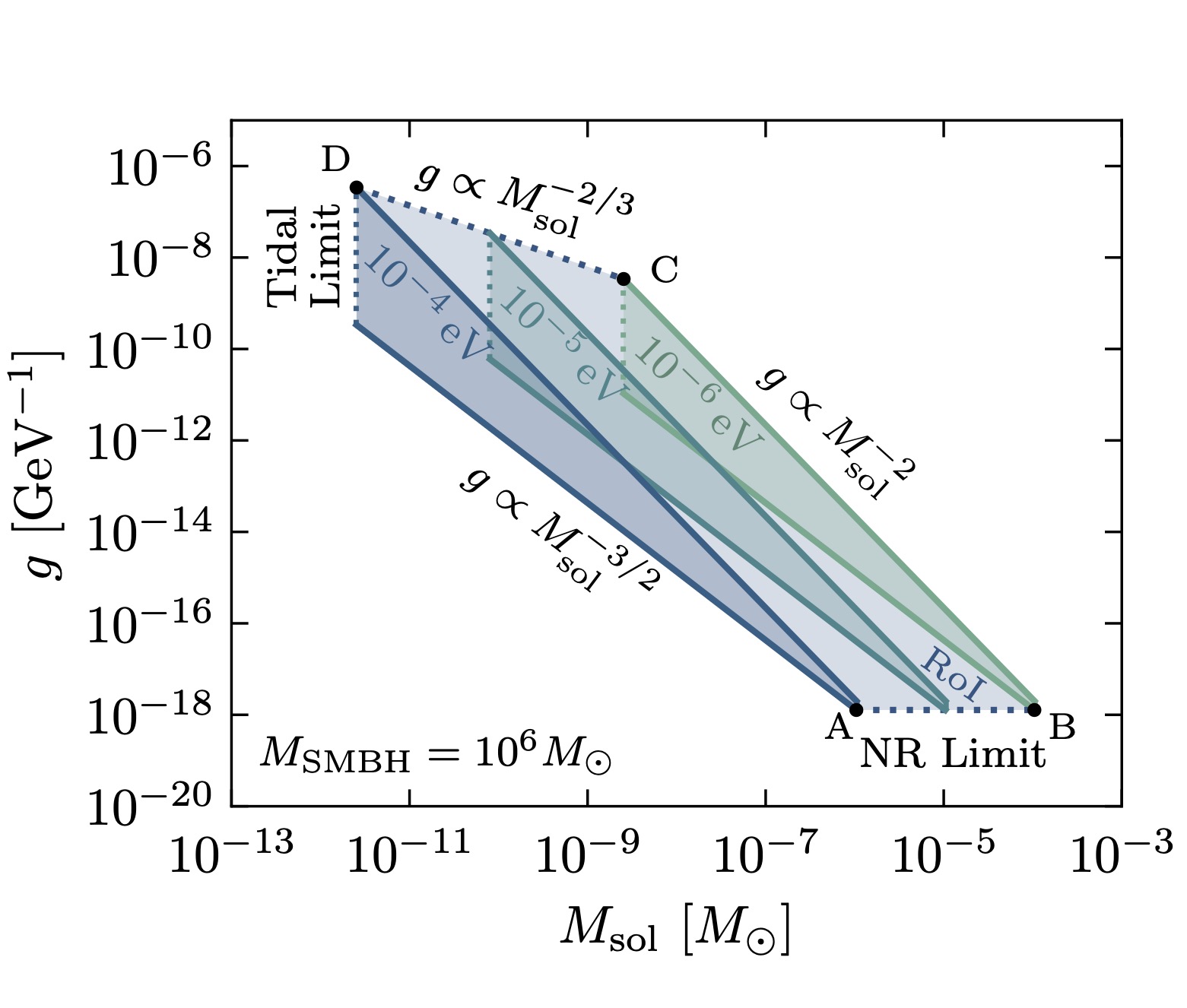}
    \end{subfigure}
    \hfill
    \begin{subfigure}[t]{0.495\textwidth}
    \includegraphics[width=0.93\textwidth]{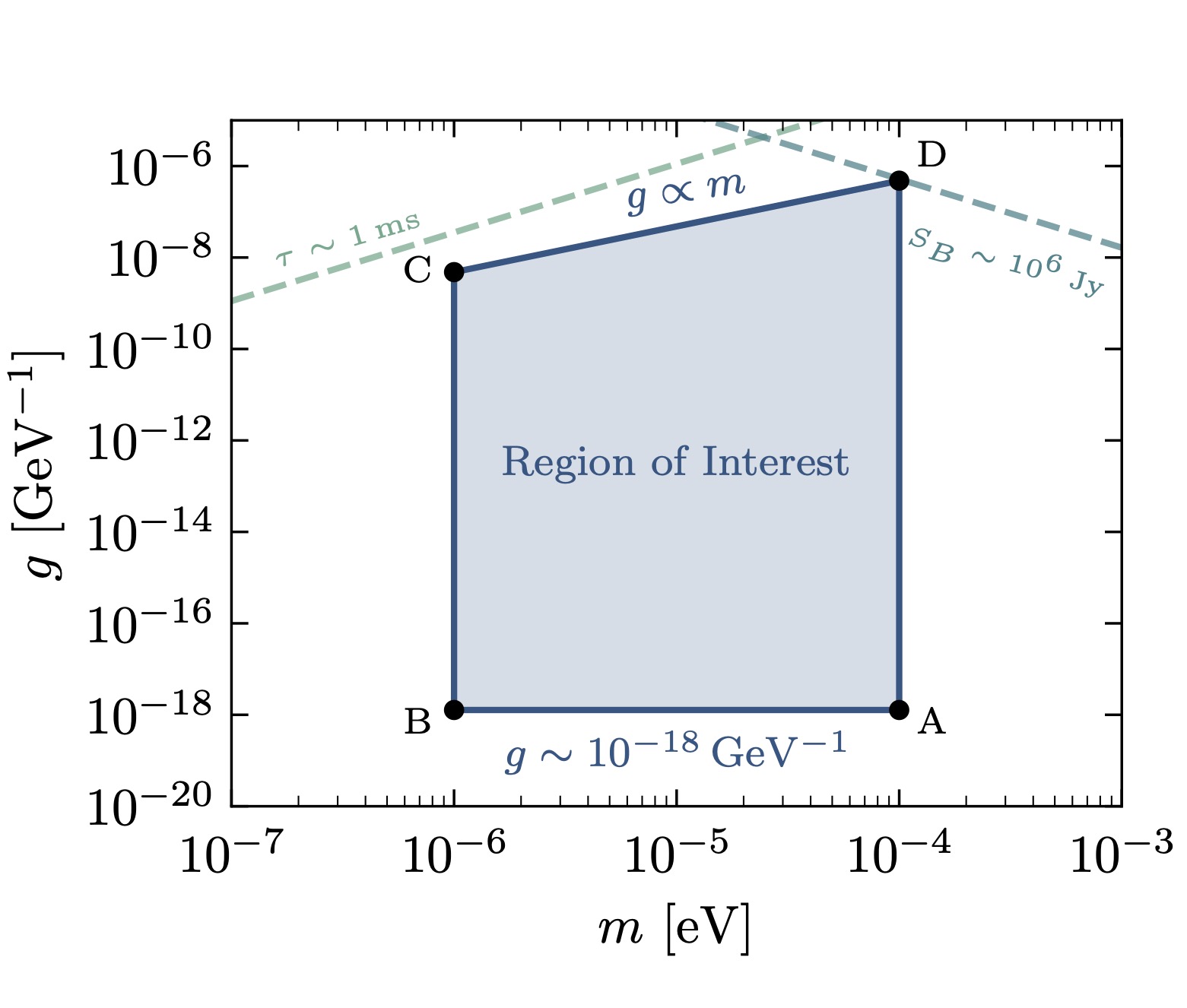}
    \end{subfigure}
    \caption{The range of the effective couplings $g$ for which both the effective field theory description holds and parametric resonance can occur. The labeled vertices---A, B, C, D---indicate how our chosen parameter spaces map onto one another. \textbf{Left:} The allowed range in $g$ with soliton mass $M_\mrm{sol}$. Each band corresponds to a given choice for the dark photon mass $m$, beginning at the tidally limited soliton masses (\cref{eq:msol_tidal}) and terminating at those for which the non-relativistic (NR) approximation becomes invalid (\cref{eq:nonrelbound}). The shaded envelope of all these bands highlights our region of interest (RoI). \textbf{Right:} The allowed range in $g$ with dark photon mass $m$. During parametric resonance, the considered range of dark photon masses gives rise to radiation observable by transient radio telescope surveys. The dashed lines show the regions of this parameter space where the burst duration $\tau$ and spectral flux density $S_B$ are similar to those of observed fast radio bursts. This is discussed in \cref{subsec:em_rad}, and their relations are given in \cref{eq:tau,eq:flux_density}, respectively.}
    \label{fig:g_regions}
\end{figure*}

\subsection{Parametric Resonance}
\label{subdsec:pr}

The parametric resonance of vector solitons has recently been proposed as a powerful tool for probing solitons composed of spin-1 dark photons~\cite{Amin:2023imi}. Combined with the oscillatory nature of the dark photon field, the coupling between this field and electromagnetism via the dimension-6 operators in \cref{eq:Operators} result in time-dependent equations of motion for the electromagnetic Fourier modes. The periodic oscillation of the soliton amplitudes then leads to the resonance phenomenon, generating radiation that carries information about the dark photon-photon coupling.

For parametric resonance to occur, the radiation must be produced faster than the rate at which it leaves the soliton, given by $\usim 2/R_\mrm{sol}$. If this condition holds, the photon occupation number then grows exponentially with a characteristic growth rate of $\usim mg^2\bar X^2/2$, leading to the efficient, Bose-enhanced emission of the ensuing radiation. A Floquet analysis shows that this condition is met if  $mg^2 \bar{X}^2/2 > 2 / R_\mrm{sol}$ \cite{Amin:2023imi}. This results in a critical mass threshold above which soliton masses must lie for parametric resonance to occur. Inserting the relations for $\bar{X}$ and $R_\mrm{sol}$ given in \cref{eq:xbar} and \cref{eq:mr_relation}, respectively, we arrive at the threshold value shown in \cref{eq:Mc}. Combining this threshold with the condition for us to remain in the non-relativistic regime, such that $g^2 \bar{X}^2 \ll 1$, we find the range of couplings for which both parametric resonance can occur and our EFT description holds:
\begin{equation}
\label{eq:g_cond}
\frac{0.552}{M_\mrm{P}} \left(\frac{m M_\mrm{sol}}{62.3 M_\mrm{P}^{2}}\right)^{-3/2} < g \ll \frac{0.490}{M_\mrm{P}} \left(\frac{m M_\mrm{sol}}{62.3 M_\mrm{P}^2} \right)^{-2} ~,
\end{equation}
where we have used the expression for the central field value $\bar{X}$ from \cref{eq:vec_sol_profile}.

Additionally, there is a third condition that the coupling strength must satisfy to ensure the tidal stability of supercritical solitons. For no significant tidal disruption around SMBHs to occur, their masses must exceed the value given in \cref{eq:msol_tidal}. Since the EFT condition gives us an upper bound on the coupling, its maximal value must then terminate at the tidally limited soliton mass. Combining the EFT requirement with that for tidal stability, we arrive at an third bound on $g$ that can be expressed in terms of either the soliton mass or the dark photon mass:
\begin{align}
g &\lesssim 7 \times 10^{-9}\,\mrm{GeV^{-1}}\left(\frac{M_\mrm{SMBH}}{10^6 M_\odot}\right)^{2/3}\left(\frac{10^{-9} M_\odot}{M_\mrm{sol}}\right)^{2/3}\,, \nonumber\\
g &\lesssim 5 \times 10^{-9}\,\mathrm{GeV}^{-1} \left( \frac{M_\mathrm{SMBH}}{10^6 M_\odot} \right) \left( \frac{m}{10^{-6}\,\mathrm{eV}} \right)
\,.
\label{eq:g_cond_tidal}
\end{align}

We illustrate the couplings that lead to parametric resonance, retain the validity of the nonrelativistic EFT, and ensure sufficient tidal stability of supercritical solitons in \cref{fig:g_regions}. We show this both with respect to the soliton mass for fixed choices of the dark photon mass and with respect to the dark photon mass itself. The shaded regions highlight our region of interest for our ensuing analysis, satisfying the conditions given in \cref{eq:g_cond,eq:g_cond_tidal}. As our ultimate merger rate calculation given in \cref{eq:MRtotal} integrates over the SMBH mass function, we fiducialize these regions to our lowest considered SMBH mass of $M_\mrm{SMBH} = 10^{6}M_\odot$ since this gives us the most constrained region of interest.

The lowest attainable value for the effective coupling strength is constrained by the requirement that we remain in the non-relativistic regime, giving $g \gtrsim 10^{-18}\,\mrm{GeV^{-1}}$ when saturating the condition given in \cref{eq:nonrelbound}. Conversely, the highest achievable couplings are limited by the lightest supercritical solitons that can be formed without significant tidal disruption around SMBHs; this is captured by \cref{eq:msol_tidal}. Since, the soliton mass itself depends on the dark photon mass, this upper bound ultimately depends on our choice for the dark photon mass.
As we will discuss in \cref{subsec:em_rad}, the dark photon masses that can produce detectable radio transient signals lies in the range $10^{-6}\,\mrm{eV} \lesssim m \lesssim 10^{-4}\,\mrm{eV}$. For a given choice of $m$, the couplings satisfying \cref{eq:g_cond} then lie within a band that begins at the lightest tidally limited soliton mass and continues until the generated $M_\mrm{sol}$ breaks the non-relativistic limit, $M_\mrm{sol} \sim 10^{-4}M_\odot \,(10^{-6}\,\mrm{eV} / m)$. Inside the highlighted regions of interest within these parameter spaces, we expect supercritical solitons to produce highly energetic bursts of radio frequency energy that are detectable by radio telescopes.

\subsection{Characteristics of Electromagnetic Radiation}
\label{subsec:em_rad}

After merging, solitons above the critical mass evaporate, producing electromagnetic radiation to which Earth-based telescopes can be sensitive. This radiation is within the radio-frequency band, with a comparatively small bandwidth. These are respectively given by~\cite{Amin:2023imi} 
\begin{align}
\nu &\sim 200\,\mrm{MHz}\,\left(\frac{m}{10^{-6}\,\mrm{eV}}\right)\,,\label{eq:nu}\\
\Delta\nu &\sim 40\,\mrm{kHz}\,\left(\frac{m}{10^{-6}\,\mrm{eV}}\right) \left(\frac{10^{-10}\,\mrm{GeV}^{-1}}{g}\right)^{2/3}\,\label{eq:delta_nu}.
\end{align}
\textcolor{black}{We note that these quantities are evaluated in the proper frame of the radiating soliton, and they can therefore incur a relativistic Doppler shift once the merged soliton velocity is accounted for. We discuss how this can influence our final results in \cref{sec:conclusion}.}

The duration of a single burst is difficult to define due to the complex dynamics that occur during the merger itself, requiring numerical simulations to capture properly. We estimate this timescale to be the inverse of the growth rate of the photon field, given by~\cite{Amin:2022pzv,Amin:2023imi}
\begin{equation}
    \label{eq:tau}
    \tau \sim  20\,\mrm{\mu s}\, \left(\frac{10^{-6}\,\mrm{eV}}{m}\right)\left(\frac{g}{10^{-10}\,\mrm{GeV^{-1}}}\right)^{2/3}\,.
\end{equation}  
However, this estimate only applies to isolated solitons and may not accurately capture the dynamics of soliton mergers. In such cases, nonlinear effects might substantially alter the emission timescale. For instance, if radiation is predominantly emitted after the merger remnant relaxes into a new stable soliton, then the relevant timescale may instead be set by the soliton crossing time, $\tau_\mrm{LC} \sim 2 R_\mrm{sol}$. 

After the merger of two solitons with masses $\usim M_c$, the mass of the final soliton is approximately given by $\usim 1.4 M_\mathrm{c}$ \cite{Hertzberg:2020dbk, Schwabe:2016rze}. Thus, the total energy released in each burst is roughly $E \sim 0.4 M_\mathrm{c}$, yielding
\begin{equation}
E \sim 1\times 10^{45}\, \mathrm{erg}\, \left( \frac{10^{-6} \,\mathrm{eV}}{m} \right) \left( \frac{10^{-10} \, \mathrm{GeV^{-1}}}{g} \right)^{2/3}\,,
\end{equation}
where $1\,\mrm{erg} \equiv 10^{-7}\,\mrm{J}$. These quantities allow us to estimate the spectral flux density and fluence of the signal:
\begin{align}
    S_B &\simeq \frac{E/\tau}{4 \pi D_\mrm{L}^2 \Delta\nu} \approx 1 \times 10^{18} \, \mathrm{Jy}\, \left(\frac{10^{-10} \, \mathrm{GeV}^{-1}}{g}\right)^{2/3} \left( \frac{10^{-6} \, \mathrm{eV}}{m} \right) \left( \frac{1 \, \mathrm{Mpc}}{D_\mrm{L}} \right)^{2}\,, \label{eq:flux_density}\\
    \mathcal{F} &\equiv S_B \tau \sim 2 \times 10^{16}\,\mrm{Jy\,ms}\,\textcolor{black}{\left( \frac{10^{-6} \, \mathrm{eV}}{m} \right)^{2}}\left( \frac{1 \, \mathrm{Mpc}}{D_\mrm{L}} \right)^{2}\,,\label{eq:fluence}
\end{align}
where we have assumed that the radiating soliton is located at a luminosity distance $D_\mrm{L}$ away and $1\,\mrm{Jy} \equiv 10^{-26}\,\mrm{W\,m^{-2}\,Hz^{-1}}$. 

We can compare the characteristics of the electromagnetic radiation emitted during soliton mergers to those of FRBs. FRBs are also brief and highly energetic bursts of radio-frequency energy, and their  astrophysical origins remain mysterious~\cite{Katz:2018xiu,Petroff:2019tty}. Thus, it is compelling to attempt to explain them using soliton mergers. However, it is difficult to reconcile all properties at once. So far, FRBs have been detected within the frequency range  $\usim 100\,\mrm{MHz}$ to $\usim10\,\mrm{GHz}$~\cite{Petroff:2016tcr,Platts:2018hiy}. Typically, FRBs have durations of $\usim 1\,\mrm{ms}$, with flux densities and fluences in the range of $5\times10^{-4}\,\mrm{Jy}$ to $100\,\mrm{Jy}$ and $5\times10^{-4}\,\mrm{Jy\,ms}$ to $100\,\mrm{Jy\,ms}$, respectively. The extremal observed values for these quantities are $\usim 10^{6}\,\mrm{Jy}$ and $\usim 10^{6}\,\mrm{Jy\,ms}$ for FRB $200428$, which is thought to have originated from within the Milky Way at a distance of $\usim 30,000\,\mrm{ly}$~\cite{CHIMEFRB:2020abu,Bochenek:2020qry}.

To compare these properties with those of our predicted signal, we must consider the range of valid parameter space in both $m$ and $g$. The dark photon masses correpsonding to the radiation frequencies above lie within the range $10^{-6}\,\mrm{eV}$ to $10^{-4}\,\mrm{eV}$.
The coupling strengths that can produce our desired signal and retain the validity of the EFT within this mass range must lie within the regions highlighted in \cref{fig:g_regions}. Within this region, we are already unable to produce $\usim1\,\mrm{ms}$ bursts: saturating our allowed parameter
space at $m \sim 10^{-6}\,\mrm{eV}$ and $g \sim 10^{-11}\,\mrm{GeV^{-1}}$ (vertex C), the longest burst duration we can produce is $\tau \sim 0.1\,\mrm{ms}$. Furthermore, we are unable to match either typical flux densities or typical fluences. For these quantities, we can best attempt to match those of FRBs by first fiducializing to our largest considered cosmic distances at $z=4$, corresponding to a luminosity distance of $D_L \sim 10\,\mrm{Gpc}$. At this distance, our closest recovered quantities are $S_B \sim 10^6\,\mrm{Jy}$---saturating the allowed parameter space at $m \sim 10^{-4}\,\mrm{eV}$ and $g \sim 10^{-6}\,\mrm{GeV^{-1}}$ (vertex D)---and \textcolor{black}{$\mathcal{F} \sim 10^5\,\mrm{Jy\,ms}$}, which are both significantly larger than the characteristic FRB values. 

Even for the extremal values from FRB 200428, whose origin is already thought to be known, \textcolor{black}{we can only match the signal fluence and flux density by extremizing our allowed region of parameter space, where this is also inconsistent with a $\usim 1\,\mrm{ms}$ duration burst.} Therefore, we claim that we cannot explain FRBs with soliton mergers. However, the non-observation of our predicted signals provides a powerful means of constraining the merger rate and, in turn, the DM fraction in vector solitons.\footnote{Solitons possessing linear or circular polarizations, corresponding to the base configurations given in \cref{eq:vec_sol_base}, can additionally generate polarized electromagnetic radiation. The operators $\mathcal{O}_3$ and $\mathcal{O}_4$ give rise to radiation that is polarized in the same manner as the radiating soliton; in contrast, the $\mathcal{O}_1$ and $\mathcal{O}_2$ operators induce unpolarized radiation for linearly polarized solitons and no radiation for circularly polarized ones. The possible polarization of the outgoing radiation is a feature that is absent for scalar DM solitons~\cite{Hertzberg:2018zte}. While we do not focus on this feature, we emphasize that a detected burst consistent with our predicted signal characteristics could be used as a smoking gun signature for the decay of supercritical vector solitons.}
\subsection{Constraining the Dark Matter Fraction}
\label{Sec:Constrain}

We can use the absence of radio bursts observed by radio telescopes consistent with the radiation predicted from vector soliton mergers to place a limit on the soliton merger rate. Radio telescopes observe an angular area of the sky $\mathcal{A}$ over a time $T_\mrm{obs}$; their total exposure is typically quoted as a product of these variables, $\varepsilon \equiv \mathcal{A} T_\mrm{obs}$. The number of events detected within this patch of the sky $N$ should follow a Poisson distribution, $N \sim \funop{\mrm{Pois}}(\lambda)$. The expectation value of this distribution $\lambda$ is then given by 
\begin{equation}
    \lambda \equiv \frac{\varepsilon \funop{\mathcal{V}(z)}}{4\pi} \,\Gamma_\mrm{merg}^\mrm{TOTAL}\,,
    \label{eq:lambda}
\end{equation}
where the total volumetric merger rate $\Gamma_\mrm{merg}^\mrm{TOTAL}$ is given in \cref{eq:MRtotal} and $\mathcal{V}(z)$ is the comoving volume at redshift $z$ within which we expect to observe the radiation from mergers. We take this to be a sphere with radius equal to the luminosity distance at $z=4$, $\mathcal{V}(z=4) = (4/3)\pi D_L^3(z=4)$. The factor of $4\pi$ in \cref{eq:lambda} accounts for expected isotropy of the incident radiation. 

To achieve a false-positive rate of $\alpha$ given no observed events, we require that $\lambda \leq  -\ln \alpha$.
For a $95\%$ confidence level limit, $\alpha = 0.05$, giving an upper limit on the expected event count of $\lambda \lesssim 3$. We may then write the upper limit on the merger rate as
\begin{align}
    \label{eq:gamma_lim}
    &\Gamma_\mrm{merg}^\mrm{TOTAL} \lesssim \Gamma_\mrm{merg}^\mrm{lim } \equiv 3\,\frac{4\pi}{\varepsilon \mathcal{V}(z)}\\
    &\simeq 2 \times 10^{-6} \,\mrm{Mpc^{-3}\,day^{-1}}\left(\frac{1\,\mrm{deg^2\,hr}}{\varepsilon}\right)\left(\frac{7332\,\mrm{Mpc}}{D_L(z)}\right)^3\,,\nonumber
\end{align}
where we have converted from steradians to square degrees to more readily incorporate the exposures of radio telescopes. 

\begin{table}[t!]
\renewcommand{\arraystretch}{1.5}
    \centering
    \begin{tabular*}{\columnwidth}{@{\extracolsep{\fill}}lcc}
    \toprule\midrule
     Telescope & \textbf{$\varepsilon~[\mrm{deg^2\,hr}]$} & $f~[\mrm{GHz}]$ \\
     \midrule
     Lorimer et al.~\cite{Lorimer:2007qn} & $4.3\times10^{3}$ & $1.23\text{--}1.52$ \\
    CHIME PF~\cite{CHIMEScientific:2017php} & $2.4 \times 10^5$ & $0.400\text{--}0.800$ \\ ASKAP~\cite{2018Natur.562..386S} & $5.1 \times 10^5$ & $1.25\text{--}1.50$ \\
     HTRU~\cite{2016MNRAS.455.2207R} & $1.7 \times 10^6$ & $1.23\text{--}1.52$\\
     CHIME (Proj.)~\cite{CHIMEFRB:2018mlh} & $5.3 \times 10^6$ & $0.400 \text{--}0.800$\\
      GBT (Proj.)~\cite{Gajjar:2018bth} & $1.0 \times 10^4$ & $4.00\text{--}8.00$\\
     \midrule
     \bottomrule      
    \end{tabular*}
    \caption{The telescopes we have considered in constraining the dark matter fraction composed of vector solitons. The relevant properties for our analysis are the exposures of these telescopes ($\varepsilon$) and the radio frequency range to which they are sensitive ($f$).}
    \label{tab:telescopes-properties}
\end{table}

We summarize the relevant details of each telescope we have considered in our analysis in \cref{tab:telescopes-properties}. We consider the flagship FRB study of Lorimer et al.~\cite{Lorimer:2007qn}, the Canadian Hydrogen Intensity
Mapping Experiment (CHIME) Pathfinder~\cite{CHIMEScientific:2017php}, the Australian Square Kilometre Array Pathfinder (ASKAP)~\cite{2018Natur.562..386S}, and the Parkes high-latitude pulsar survey (HTRU)~\cite{2016MNRAS.455.2207R}. We also consider projections with CHIME and the Green Bank Telescope (GBT)~\cite{Gajjar:2018bth}. 

For our projections, we assume a total project lifetime of three years and a total covered angular sky area of $200\,\mrm{deg^2}$ in the case of CHIME~\cite{CHIMEFRB:2018mlh}. For the Green Bank Telescope, we assume a total exposure of $10^{4}\,\mrm{deg^2\,hr}$ to showcase the sensitivity that a high-exposure version of a GBT-like telescope can achieve. Typically, these high frequency telescopes conduct targeted searches, focusing on a particular FRB source and therefore admitting much lower exposures than the ones possessed by lower frequency telescopes. Nonetheless, taking a total sky exposure per beam of $\theta^2 \sim (\lambda / D)^2 \sim 5 \times 10^{-4}\,\mrm{deg^2}$ for a frequency of $f = 8\,\mrm{GHz}$ and the GBT dish diameter of $D \sim 100\,\mrm{m}$, we can achieve this higher exposure using, for instance, a dedicated array of approximately $20$ telescopes each operating with $36$ beams over three years. These specifications are similar to those planned for ASKAP; however, we emphasize that we consider this projection only to highlight the difference in sensitivity in our parameter spaces for telescopes operating at higher radio frequencies.

\begin{table}[t!]
\renewcommand{\arraystretch}{1.5}
\setlength{\tabcolsep}{4pt}
    \centering
    \small
    \begin{tabular*}{\textwidth}{@{\extracolsep{\fill}}p{3.25cm}ccccc}
    \toprule\midrule
     Telescope &  \textbf{$\Gamma^\mrm{lim}_\mrm{merg}$} & $f_\mrm{DM}^\mrm{min}$ & $M_\mrm{sol}$ & $m$ & $g_\mrm{max}$\\
      &  \textbf{$[\mrm{Mpc^{-3}\,day^{-1}}]$} &  & $[M_\odot]$ & $\mrm{[eV]}$ & $\mrm{[GeV^{-1}]}$\\
     \midrule
     Lorimer et al.~\cite{Lorimer:2007qn} & $4.2 \times 10^{-10}$ & $7.0\times 10^{-2}$ & $1 \times10^{-10}\text{--}2\times10^{-5}$& $(5.1\text{--}6.3) \times 10^{-6}$ & $3.0 \times10^{-8}$   \\
    CHIME PF~\cite{CHIMEScientific:2017php} & $7.5 \times 10^{-12}$ & $9.4\times 10^{-3}$ & $4 \times 10^{-10}\text{--}6\times 10^{-5}$ & $(1.7\text{--}3.3) \times 10^{-6}$ & $1.6 \times10^{-8}$  \\
     ASKAP~\cite{2018Natur.562..386S} & $3.5 \times 10^{-12}$ & $6.4\times 10^{-3}$ & $2 \times 10^{-10}\text{--}2\times 10^{-5}$ & $(5.2\text{--}6.2) \times 10^{-6}$ & $3.0 \times10^{-8}$ \\
     HTRU~\cite{2016MNRAS.455.2207R} & $1.1 \times 10^{-12}$ & $3.5\times 10^{-3}$ & $2 \times 10^{-10}\text{--}2\times 10^{-5}$ &$(5.1\text{--}6.3)\times10^{-6}$ & $3.0 \times10^{-8}$ \\
     CHIME (Proj.)~\cite{CHIMEFRB:2018mlh} &  $3.4 \times 10^{-13}$ & $2.0\times 10^{-3}$ & $4 \times 10^{-10}\text{--}6\times 10^{-5}$ & $(1.7\text{--}3.3) \times 10^{-6}$ & $1.6 \times10^{-8}$ \\
      GBT (Proj.)~\cite{Gajjar:2018bth} & $1.8 \times 10^{-10}$ & $4.6\times 10^{-2}$ & $1 \times 10^{-11}\text{--}6\times 10^{-6}$ & $(1.7\text{--}3.3) \times 10^{-5}$ & $1.6 \times10^{-7}$ \\
     \midrule
     \bottomrule    
    \end{tabular*}
    \caption{The inferred properties from our analysis for the telescopes we have considered. These include the upper limit on the soliton merger rate ($\Gamma^\mrm{lim}_\mrm{merg}$), the minimum constraint on the dark matter fraction in vector solitons using our mean $95\%$ confidence level limit ($f_\mrm{DM}^\mrm{lim}$), and the range of soliton masses ($M_\mrm{sol}$); dark photon masses ($m$); and maximal couplings ($g_\mrm{max}$) probed by each telescope. The minimum coupling is always $g_\mrm{min} \approx 1.3 \times 10^{-18}\,\mrm{GeV^{-1}}$. The relevant details for our analysis for each telescope are summarized in \cref{tab:telescopes-properties} and our results are visualized in \cref{fig:g_regions}.}
    \label{tab:telescopes-inference}
\end{table}

We show our results for $\Gamma_\mrm{merg}^\mrm{lim}$ in \cref{tab:telescopes-inference}. All of our inferred rates are consistent with being lower than the measured FRB rate, standing at $\Gamma_\mrm{FRB} \sim 10^{-9}\,\mrm{Mpc^{-3}\,day^{-1}}$~\cite{Petroff:2019tty}. We see that we achieve greater sensitivities with higher telescope exposures, as also indicated by \cref{eq:gamma_lim}. Our weakest and strongest limits respectively arise from the first FRB discovery study by Lorimer et al.~\cite{Lorimer:2007qn} and the HTRU survey~\cite{2016MNRAS.455.2207R}, which yield $\Gamma_\mrm{merg}^\mrm{TOTAL} \lesssim 4.2 \times 10^{-10}\,\mrm{Mpc^{-3}\,day^{-1}}$ and $\Gamma_\mrm{merg}^\mrm{TOTAL} \lesssim 1.1 \times 10^{-12}\,\mrm{Mpc^{-3}\,day^{-1}}$. Our best projected limit follows from the full CHIME anticipated exposure~\cite{CHIMEFRB:2018mlh}, giving $\Gamma_\mrm{merg}^\mrm{TOTAL} \lesssim 3.4 \times 10^{-13}\,\mrm{Mpc^{-3}\,day^{-1}}$.

We can convert our upper limits on $\Gamma_\mrm{merg}^\mrm{lim}$ to ones on the DM fraction that vector solitons can compose $f_\mrm{DM}$. Using our predictions of the merger rate given in \cref{fig:MR}, which assume a DM fraction of unity, and the fact that $\Gamma_\mrm{merg}^\mrm{TOTAL} \propto f_\mrm{DM}^2$ from \cref{eq:merger}, we then have that
\begin{equation}
\label{eq:fDMlimit}
f_\mrm{DM}(M_\mrm{sol}) \lesssim \sqrt{\frac{\Gamma^\mrm{lim}_\mrm{merg}}{\Gamma_\mrm{merg}^\mrm{TOTAL}(M_\mrm{sol};f_\mrm{DM} = 1)}}\,.
\end{equation}
Thus, for a given telescope, we can identify an upper limit on the fraction of DM fraction that solitons can compose given the non-observation of our predicted signal by radio-transient surveys.

To compute this limit, we first convert the frequency range to which a telescope is sensitive to the dark photon masses that would produce these same radiation frequencies using $m = 2 \pi f$. For the maximum and minimum dark photon masses, corresponding to the maximum and minimum of a telescope's frequency range, we then find the range of soliton masses that retain tidal stability and conform to the non-relativistic approximation via \cref{eq:msol_tidal,eq:nonrelbound}, respectively. This yields two bands within which our upper limit can lie that incorporate the uncertainties in the SMBH mass function, shown in \cref{fig:PhiSMBH}. We then take the envelope of these bands to represent the region where the true limit should lie. Finally, we take the geometric mean of the limits at the boundaries of this envelope to be the best estimator of our limit. For all of our limits and projections, we assume the density profile index  $\gamma =1$, showing the limits for other $\gamma$ values in \cref{sec:app-all_gamma}. Since higher $\gamma$ values lead to larger merger rates, this choice gives us the most conservative upper limits.

\begin{figure*}[t]
    \centering
    \begin{subfigure}[t]{0.495\textwidth}
    \includegraphics[width=0.93\textwidth]{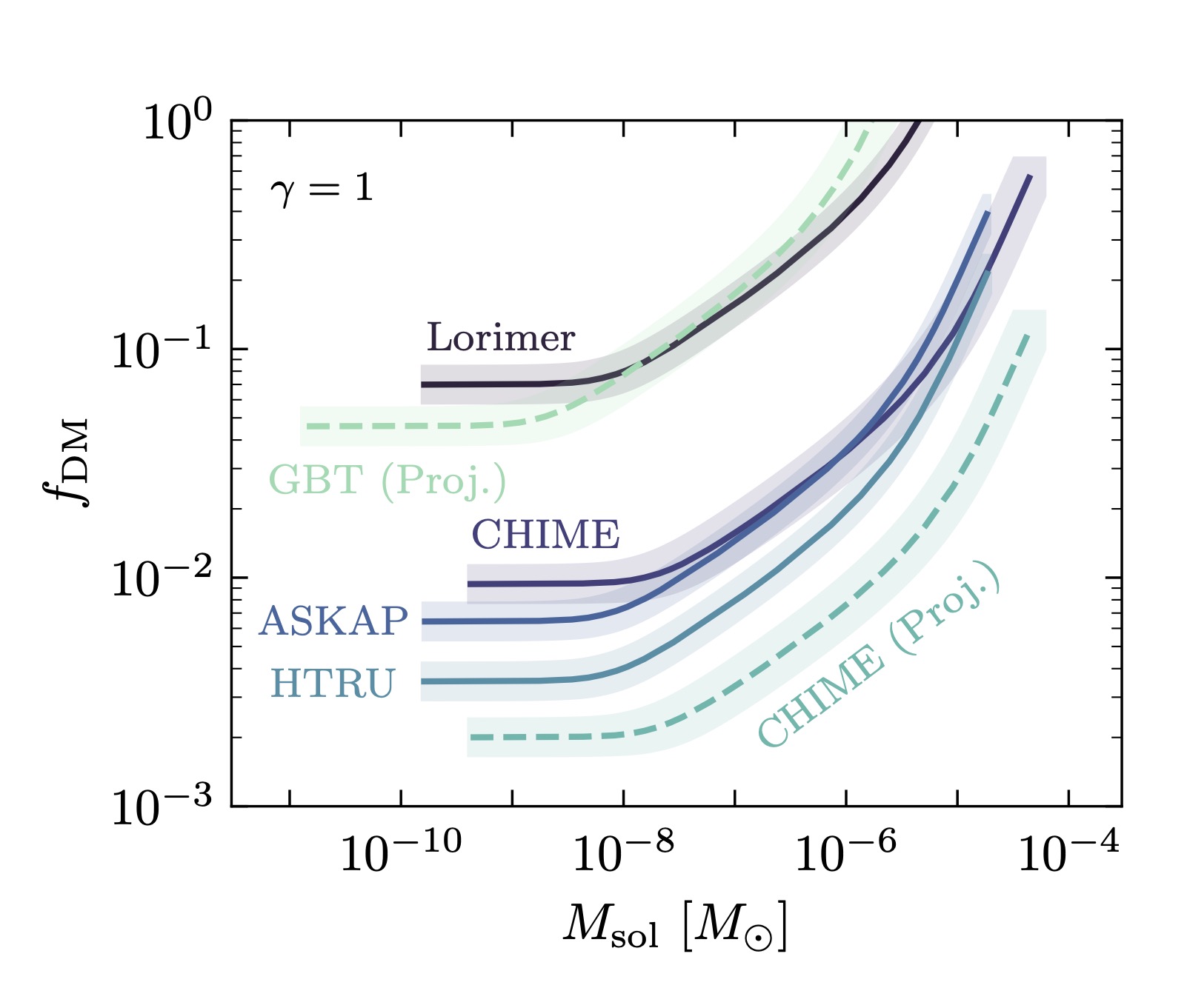}
    \end{subfigure}
    \hfill
    \begin{subfigure}[t]{0.495\textwidth}
    \includegraphics[width=0.93\textwidth]{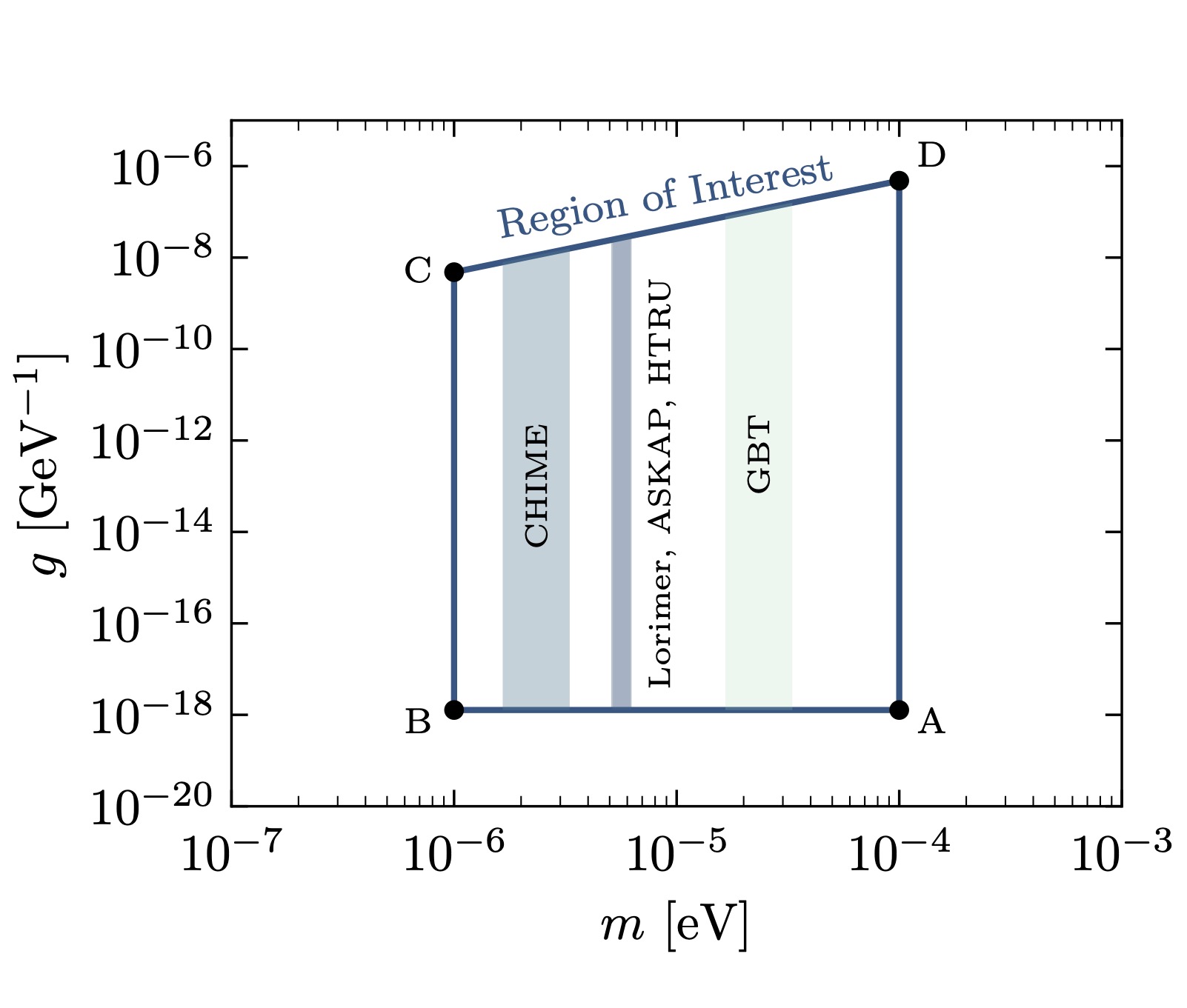}
    \end{subfigure}
    \caption{\textbf{Left:} The $95\%$ confidence level upper limits on the fraction of dark matter that vector solitons can compose $f_\mrm{DM}$ with soliton mass $M_\mrm{sol}$. Limits and projections are shown for the telescopes considered in \cref{tab:telescopes-properties}. The bands correspond to the where each limit can lie as a result of the uncertainty in the merger rate, shown in \cref{fig:MR}, and the range of frequencies probed by each telescope. Each band commences at the minimum soliton mass required for tidal stability around supermassive black holes (\cref{eq:msol_tidal}) and terminates when the non-relativistic (NR) approximation becomes invalid (\cref{eq:nonrelbound}). The solid lines show the geometric mean of the limits at the boundaries of each band, which we take to be the best estimator of our limits. Results are shown for the density profile index $\gamma = 1$; we show the results for all of our considered indices in \cref{sec:app-all_gamma}.  \textbf{Right:} The region of the parameter space of the effective coupling $g$ and dark photon mass $m$ probed by our considered telescopes within the region of validity of our analysis (\cref{fig:g_regions}). Numerical results are summarized in \cref{tab:telescopes-inference}.}
    \label{fig:fdm_lim}
\end{figure*}

We show our $95\%$ confidence level upper limits on the fraction of DM that vector solitons can compose for each of our considered telescopes in \cref{fig:fdm_lim}. Since each telescope has a characteristic frequency range of sensitivity, the dark photon masses they are able to probe varies, leading to distinct ranges for the probable soliton masses. Generally, the higher the probed frequency is, the lower the soliton mass that a telescope can detect. We also show the regions of the parameter space in the effective coupling $g$ and dark photon mass $m$ to which each telescope is sensitive. We embed these parts of the parameter space within the complete region of interest for our analysis, shown in \cref{fig:g_regions} and explained further in \cref{subdsec:pr}.

We summarize our main results in \cref{tab:telescopes-inference}, which includes the minimum DM fraction constrained $f_\mrm{DM}^\mrm{min}$, the range of probable soliton masses, the dark photon masses probed, and the maximum effective coupling strength reached $g_\mrm{max}$. The most stringent constraints on $f_\mrm{DM}$ occur at the lowest soliton masses, where the merger rate is largest, as shown in \cref{fig:MR}. From the first FRB study by Lorimer et al.~alone, we can already constrain the DM fraction in vector solitons to be $f_\mrm{DM} \lesssim 7.0 \times 10^{-2}$. CHIME Pathfinder, ASKAP, and the Parkes HTRU survey are all approximately one order of magnitude more constraining, such that $f_\mrm{DM}$ can be at most $\lesssim 9.4 \times 10^{-3}$, $\lesssim 6.4\times 10^{-3}$, and $\lesssim 3.5 \times 10^{-3}$, respectively. Our projection for the full CHIME exposure yields the strongest constraint, with $f_\mrm{DM} \lesssim 2.0 \times 10^{-3}$, whereas our GBT projection, though less constraining, probes the lowest soliton mass of $M_\mrm{sol} \approx 1 \times 10^{-11}M_\odot$.

Within the parameter space of the effective coupling $g$ and dark photon mass $m$, each telescope is sensitive to a specific region dictated by its probable frequency range. Given the non-observation of a radio signal consistent with our predictions, a telescope can exclude its relevant part of the parameter space if the DM fraction in vector solitons $f_\mrm{DM}$ lies above the value that the telescope can constrain. If instead $f_\mrm{DM}$ is lower than this upper limit, the band is once again allowed. Outside of any given region, a telescope is insensitive to the parameter space as either it cannot observe the frequency of the emitted radiation due to a merger, the supercritical soliton created by a merger is significantly tidally disrupted such that parametric resonance does not appreciably occur, or the EFT framework on which our analysis is based is rendered invalid due to the breakdown of the non-relativistic approximation. 

Owing to its low frequency range, CHIME is able to constrain the smallest dark photon masses and couplings. The study by Lorimer et al., ASKAP, and the Parkes HTRU survey instead probe an intermediate part of the parameter space over a thin band due to their smaller frequency range. Finally, our hypothetical, high-exposure realization of GBT is sensitive to the highest dark photon masses and couplings thanks to its high frequency range. To completely cover our region of interest, telescopes would have to probe the frequencies $100\,\mrm{MHz}$ to $10\,\mrm{GHz}$; however, as discussed above, high frequency signals are challenging to constrain due to the difficulty in achieving large exposures.

Furthermore, to detect such radiation using terrestrial radio telescopes, the mass of the spin-$1$ particle must be at least \( \usim10^{-7}\,\text{eV} \). The lower limit is associated with a signal frequency of approximately \( 30 \text{ MHz} \); detecting frequencies below this limit is particularly difficult due to the ionosphere's absorption and scattering of low-frequency photons. Considering instead future space-based facilities, the minimum mass limit could decrease to lower values. For instance, the Orbiting Low Frequency Antennas for Radio Astronomy Mission (OLFAR)~\cite{2020AdSpR..65..856B, 2016ExA....41..271R} plans to launch thousands of nanosatellites on the far side of the Moon, allowing us to detect signals with frequencies as low as \( 0.30\, \text{MHz} \). OLFAR would therefore lower the minimum mass of spin-$1$ particle to 
$\usim 10^{-9}\,\text{eV}$.\footnote{Monitoring from space has been explored in different scenarios pertaining to indirect investigations of DM. As an illustration, see Ref.~\cite{Choi:2022btl} regarding the identification of ultra-long radio waves resulting from axion-photon transformation during encounters between axion self-similar minihalos and neutron stars.}

While FRB surveys have not yet observed our predicted signal, the radiation produced from the decay of vector solitons would leave a unique signature for radio telescopes. Not only would it be distinct from those cataloged as fast radio bursts, but it would also differ from that of the decay of scalar solitons~\cite{Hertzberg:2018zte}. Due to the scalar (or pseudo-scalar) nature of a hypothetical light bosonic spin-$0$ DM particle, the radiation output from scalar solitons is expected to be isotropic. In contrast, depending on the coupling between dark photons and photons and the polarization state of the solitons, the produced radiation from vector solitons can exhibit a specific polarization pattern~\cite{Amin:2023imi}. Thus, in the event of a detection, radio telescopes would readily be able to identify our signal with the decay of vector solitons via parametric resonance, providing evidence for the existence of DM and dark photon solitons.   

\section{Discussion and Conclusions}
\label{sec:conclusion}

We have presented the first constraints on the DM fraction in spin-$1$ dark photon solitons $f_\mrm{DM}$ by considering their mergers in the dense astrophysical environments surrounding SMBHs. We began by providing a novel computation of the soliton merger rate, accounting for the spiky DM density profiles that arise from the adiabatic growth of SMBHs at galactic centers. By incorporating the effects of the enhanced soliton densities and velocity dispersions in these regions, we computed the total merger rate to be $\Gamma_{\text{merg}}^{\text{TOTAL}} \lesssim 10^{-7}f^2_{\text{DM}}\,\text{Mpc}^{-3}\,\text{day}^{-1}$ for an initial DM profile $\rho_\mrm{DM} \propto r^{-1}$, with steeper fall offs resulting in higher rates. We found this rate to be dependent on the soliton mass, which ranged from the minimum mass required for tidal stability around SMBHs to the maximum mass retaining the validity of our EFT approach.

Since vector solitons are coherent oscillating compact objects, they can undergo parametric resonance of photons for sufficiently strong dark photon-photon couplings or, equivalently, for sufficiently massive solitons. Considering solitons with masses just below this threshold, soliton mergers provide a mechanism for obtaining supercritical solitons and, consequently, for producing the highly energetic bursts of radio-frequency radiation via parametric resonance that are readily observable by radio telescopes.

Equating our computed rate to that at which supercritical solitons form, we then used the non-observation of the radio transient signals produced from their decay to statistically constrain DM properties.
Combining the non-observation of such signals with our theoretical predictions of both the merger rate and the signal characteristics, we first derived upper bounds on $f_\mrm{DM}$. We considered the sky exposures from the Parkes telescope used in the first fast radio burst study by Lorimer et al., CHIME, ASKAP, the Parkes HTRU survey, and a high-exposure realization of the Green Bank telescope. We found that the first fast radio burst study alone already limits the soliton fraction to $f_\mrm{DM} \lesssim 7\times10^{-2}$, with HTRU tightening this to $f_\mrm{DM} \lesssim 4 \times 10^{-3}$, and CHIME projected to reach $f_\mrm{DM} \lesssim 2 \times 10^{-3}$ at full exposure.

For DM fractions larger than our constraints, we instead excluded parts of the parameter space in the dark photon mass $m$ and the effective dark photon-photon coupling $g$ associated with dimension-$6$ operators. Within these telescope-specific regions of the parameter space, the produced radiation from the solitonic decay would be consistent with a detectable signal by our considered telescopes. These regions were dependent on the operable frequency range for a given telescope, controlling the sensitivity to the dark photon mass. We found that CHIME can exclude the lowest dark photon masses at $m \sim 10^{-6}\,\mrm{eV}$, with our high-exposure realization of GBT sensitive to the highest masses at $m \sim 10^{-5}\,\mrm{eV}$.  Retaining the validity of our non-relativistic EFT approach and ensuring sufficient tidal stability of vector solitons around SMBHs, we excluded effective couplings in the wide range of $10^{-18}\,\mrm{GeV^{-1}} \lesssim g \lesssim 10^{-8}\,\mrm{GeV^{-1}}$.

Our results have been derived relying on a set of assumptions, which we summarize below; we include further details in \cref{sec:assumptions}. We have firstly assumed that supermassive black holes form at the centers of galaxies adiabatically, leading to the formation of spiky dark matter density profiles. Relaxing adiabaticity and
centrality, as well as including additional dynamical effects, can flatten,or in some circumstances sharpen,
the resulting spike. Secondly, we have accounted for the uncertainties in the supermassive black hole mass function at varying redshift values, which translates to a band in both the soliton merger rate and, therefore, our limits on the dark matter fraction. Thirdly, we have neglected the effect of the relativistic Doppler shift on the characteristics of the observed frequency. The impact of this is to at most shift our dark photon mass sensitivities by $\usim40\%$, as we explicitly show in \cref{subsec:doppler}. Fourthly, we have ignored the effect of the interstellar medium in attenuating our generated signal---this serves to relax our bounds but only in pessimistic cases where the radio-frequency radiation passes through highly optically thick warm/dense regions, as shown in \cref{subsec:radiationescape}. Finally, we have taken a monochromatic mass function for solitons to focus on the fraction of solitons at a given mass. This was to remain general and model independent since the details of the mass function depend on the formation mechanism; nonetheless, we include a toy calculation in \cref{subsec:solitonnonmono}, showing that our calculation is sensitive to the choice of production mechanism.

Our findings open a new direction into the phenomenology of ultralight vector fields, demonstrating how observational constraints on soliton dynamics can inform and refine theoretical models of dark matter. More broadly, this work highlights the power of \textit{radio silence} as a phenomenological probe. The evolution, interactions, and eventual fates of compact dark objects leave testable imprints in the transient sky, opening the door to future multi-messenger searches that may finally reveal the nature of dark matter.

\begin{acknowledgments}
We would like to thank Mustafa A.~Amin, Andrew J. Long, and Paola Arias for enriching discussions. EDS acknowledges support from the FONDECYT project N$^{\text{o}}$ 1251141 (Agencia Nacional de Investigaci\'on y Desarrollo, Chile). DA was jointly supported by NSF Award 2046549 and ERC grant ERC-2024-SYG 101167211 by the European Union. HYZ acknowledges support from the Shanghai Magnolia Plan Pujiang Program 25PJA073, CPSF General Program 2025M783412, and NSFC Grant No. 12350610240. Views and opinions expressed are however those of the author(s) only and do not necessarily reflect those of the European Union, European Research Council Executive Agency, or other awarding body. Neither the European Union nor the granting authority can be held responsible for them.
\end{acknowledgments}


\appendix

\section{Soliton Merger Rate in Milky-Way-like Galaxies}
\label{App:MRMW}

The rate at which solitons merge in a galaxy per unit time for two solitons each with mass \( M_{\text{sol}} \) can be expressed as,
\begin{equation}
\Gamma_{\text{merg}} =  \int^{r_\text{vir}}_{2 r_\mrm{Sch}}\frac{4\pi r^2}{2} 
\langle \sigma_{\text{eff}}(v_{\text{rel}})v_{\text{rel}}  \rangle_{\text{merg}} \left(  \frac{\rho_{\text{DM}}(r)f_{\text{DM}}}{M_{\text{sol}}} \right)^2  \dd r\,,
\label{eq:MR}
\end{equation}
where $r$ is the galactic radius, $f_{\text{DM}}$ represents the fraction of vector solitons that make up DM, the galactic DM density profile within the galactic DM halo with virial radius $r_\text{vir}$ is given by $\rho_{\text{DM}}(r)$,  $v_{\text{rel}}$ is the relative velocity between vector solitons, $\sigma_{\text{eff}}$ is the effective cross section, and $\langle \sigma_{\text{eff}}(v_{\text{rel}})v_{\text{rel}} \rangle_{\text{merg}}$ is the effective cross section averaged over the soliton relative velocity. The integration lower bound is twice the Schwarzschild radius $2 r_\mrm{Sch} =4 G_\mrm{N} M_\mathrm{SMBH}$, where we take the mass of the SMBH to be $M_{\text{SMBH}} \approx 3.5 \times 10^6M_{\odot}$, corresponding to that of Sagittarius A*\,\cite{2005ApJ...628..246E, 2005ApJ...620..744G, Ghez:2003rt, 2003ApJ...596.1015S}. The factor of $1/2$ is included to avoid double-counting since mergers occur between the same kind of objects. The effective merger cross section averaged over the soliton relative velocity reads as~\cite{Schiappacasse:2025mao} 
\begin{align}
\langle \sigma_{\mathrm{eff}}\,v_{\mathrm{rel}}  \rangle_{\mathrm{merg}} =
\int_0^{\mathrm{min}\left(2v_{\mathrm{esc}},v^*_{\mathrm{rel}}\right)}\funop{p({v_{\mathrm{rel}}})} \sigma_{\mathrm{eff}}\,v_{\mathrm{rel}}\dd v_{\mathrm{rel}}\,,
\label{eq:averagevrel}
\end{align}
where $v^*_{\mathrm{rel}}$ saturates the condition in \cref{eq:mergercond}. The upper integration limit ensures that
collisions lead to mergers. 

The effective cross section reads
\begin{equation}
\begin{split}
&\sigma_{\text{eff}}(v_{\text{rel}}) = 4\pi R^2_{\text{sol}} \left( 1 +   \frac{M_{\text{sol}}}{4\pi M_\mrm{P}^2 R_{\text{sol}}v^2_{\text{rel}}} \right)\,, 
\\  
&\simeq
5 \times 10^{5}\,\text{km}^2\, \left(\frac{10^{-9}M_{\odot}}{M_{\text{sol}}}\right)^2\left(\frac{10^{-6}\,\text{eV}}{m}\right)^4 
\Bigg\{1 +10^{-5} \left[
\left(\frac{m}{10^{-6}\,\text{eV}}\right)^2\left(\frac{M_{\text{sol}}}{10^{-9}\,M_{\odot}}\right)^2
\left(\frac{\sqrt{2}\times 220\,\text{km/s}}{v_{\text{rel}}}\right)^2 \right]\Bigg\}\,,
\end{split}
\label{sigmaeff}
\end{equation}
with $v_{\text{esc}} \equiv (2G_\mrm{N} M_{\text{gal}}(r)/ r)^{1/2}$ the soliton escape velocity. The quantity $M_{\text{gal}}(r)$ is the total galactic mass contained within the radius $r$, and $p(v_\mrm{rel})$ is the probability density function governing the relative velocities of solitons with dark galactic halos. We take this to be a 3D isotropic Maxwell-Boltzmann distribution, such that
\begin{equation}
\funop{p({v_{\text{rel}}})}\mrm{d}v_{\text{rel}} = 4 \pi v^2_{\text{rel}}\left( \frac{3}{2\pi\sigma^2_{\text{rel}}} \right)^{3/2}  
 \exp\left(-\frac{3v^2_{\text{rel}}}{2\sigma^2_{\text{rel}}}\right) \dd v_{\text{rel}}\,,
\label{eq:Pvrel}
\end{equation}
where $\sigma_{\text{rel}}$ is the 3D relative velocity dispersion for soliton encounters. This quantity varies across different regions of the galaxy. 

To estimate $\sigma_{\text{rel}}$ in the outer galactic regions, we use the Eddington inversion formula in conjunction with a series of Monte Carlo simulations. We then fit the numerically obtained relative velocity distribution functions to the distribution in \cref{eq:Pvrel}. 
Conversely, in the inner galactic regions, we use the spherical Jeans equation. For this, we assume that soliton velocities are distributed according to \cref{eq:Pvrel} and that the DM density profile follows a Navarro-Frenk-White (NFW) profile for at these smaller radii~\cite{1996ApJ...462..563N}.  
Assuming that the dispersion is isotropic, such that
$\sigma_r(r) = \sigma_{\theta}(r)= \sigma_{\phi}(r)\equiv \sigma_{\text{DM}}(r)$, we may then write the Jeans equation as
\begin{equation}
\frac{\funop{\partial(\rho_{\text{DM}}(r)}\sigma_{\text{DM}}^2)}{\partial r} + \funop{\rho_{\text{DM}}(r)}\frac{\partial \Phi_\mrm{N}(r)}{\partial r} =0 \,,  
\end{equation}
where the gravitational potential at radius $r$ is given by $\Phi_\mrm{N}(r) \equiv -G_\mrm{N} M_{\text{SMBH}}/r$. The solution of this differential equation is
\begin{equation}
\sigma_{\text{DM}}(r) = \sqrt{ \frac{G_\mrm{N} M_{\text{SMBH}}}{2r}}\,,
\label{eq:Jeq}
\end{equation}
which is valid for $r \lesssim r_\mrm{h}$, where $r_\mrm{h}$ is the gravitational influence radius around Sagittarius A*. This is given by $r_\mrm{h}  \equiv G_\mrm{N} M_{\text{SMBH}}/\sigma_{*}^2$\,\cite{1972ApJ...178..371P}, with $\sigma_{*} \sim 85\,\mrm{km\,s^{-1}}$ as the stellar velocity dispersion of the host bulge (using the mass-velocity dispersion relation given in Ref.~in~\cite{2000ApJ...539L...9F}), and $\sigma_{\text{rel}} \equiv \sqrt{2}\sigma_{\text{DM}}$.
Using the BCKM-model \cite{Bhattacharjee:2012xm} for the Milky Way Galaxy as a benchmark model for a spiral galaxy, we reproduce the previous result from  Ref.~\cite{Schiappacasse:2025mao}.

\begin{figure}
    \centering
    \includegraphics[width=0.93\textwidth]{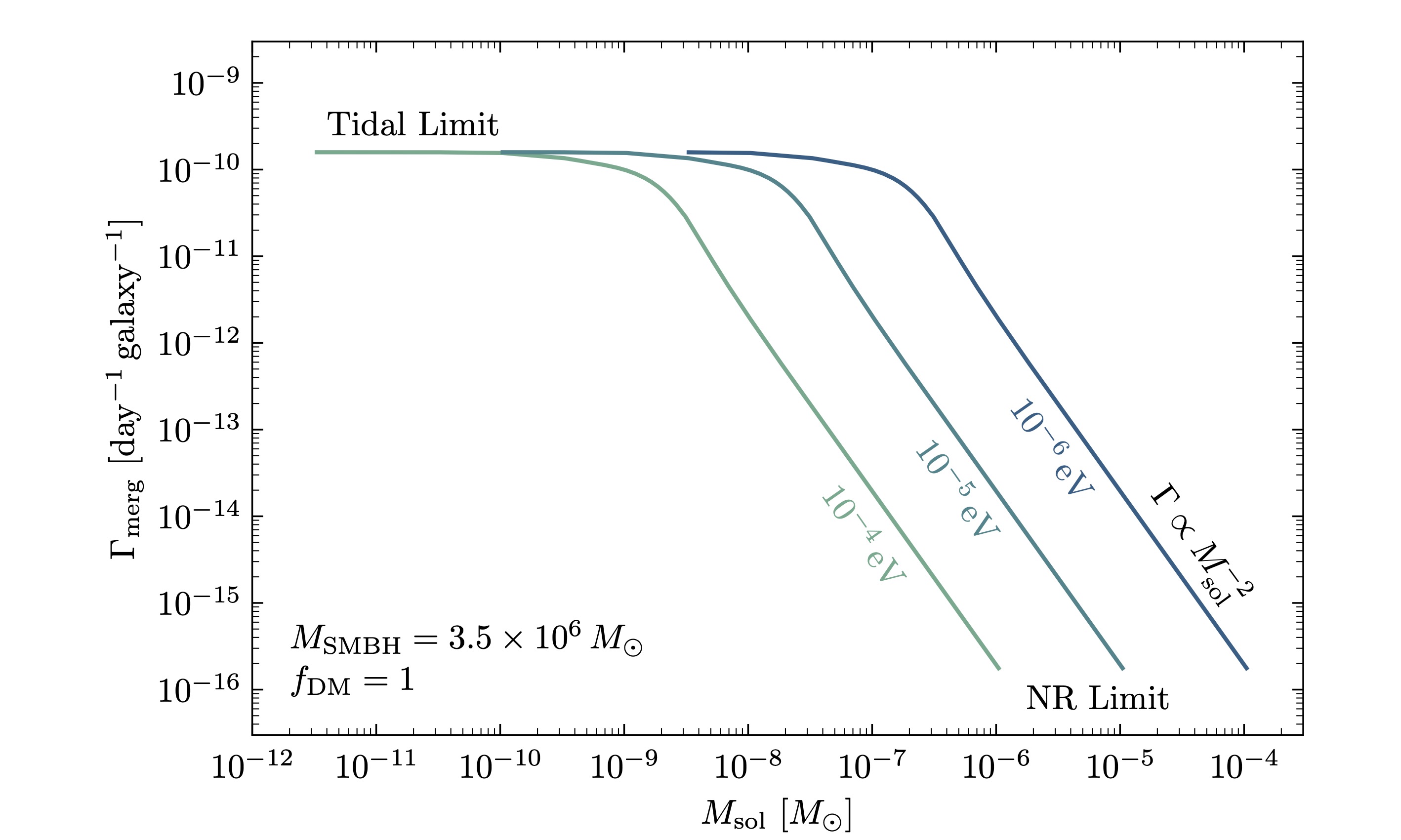}
    \caption{\justifying The soliton merger rate for Milky-Way-like galaxies $\Gamma_\mrm{merg}$ with soliton mass $M_\mrm{sol}$ in the presence of a supermassive black hole of mass equal to that of Sagittarius A$^*$, $M_\mrm{SMBH} = 3.5 \times 10^{6}M_\odot$. We assume that vector solitons compose all of the dark matter, such that $f_\mrm{DM} = 1$. The accessible soliton masses are limited from below by the constraint that they must remain tidally stable around the supermassive black hole (\cref{eq:msol_tidal}) and from above by the condition that we must remain in the non-relativistic (NR) regime (\cref{eq:nonrelbound}).}
    \label{fig:MR1}
\end{figure}

\begin{figure}
    \centering
    \includegraphics[width=0.93\textwidth]{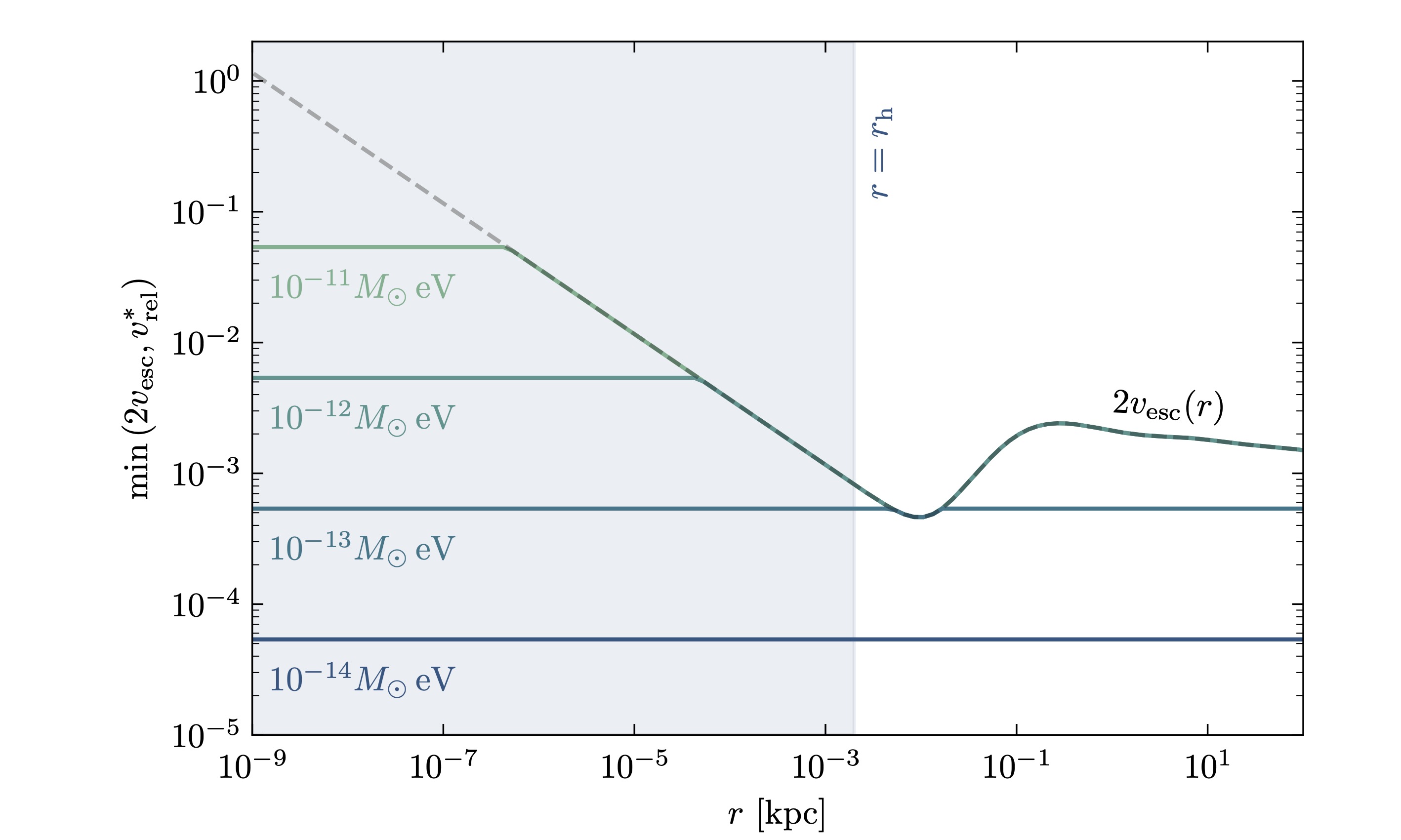}
    \caption{\justifying The integration upper limit of \cref{eq:averagevrel} with respect to the galactic radius $r$. The solid lines show the value of the upper limit labeled according to the assumed value of $M_\mrm{sol} m$, where $M_\mrm{sol}$ is the soliton mass and $m$ is the dark photon mass. The dashed black line indicates the value of twice the escape velocity for a soliton orbiting at radius $r$, with $v_\mrm{esc}(r) \equiv (2 G_\mrm{N} M_\mrm{gal}(r)/r)^{1/2}$. At small galactic radii, the upper limit is dictated by the value of the relative velocity saturating \cref{eq:mergercond} $v_\mrm{rel}^*$ for high values of $M_\mrm{sol} m$. The escape velocity instead becomes important at larger radii. The shaded region indicates where our analysis is valid, $r \leq r_\mrm{h} \approx 2\,\mrm{pc}$, with $r_\mrm{h}$ the gravitational influence radius.}
    \label{fig:MR2}
\end{figure}

We show the number of dark photon soliton mergers per day in a Milky-Way-like galaxy with respect to the soliton mass in \cref{fig:MR1}. We show this for different dark photon masses $m$ assuming $f_{\text{DM}}=1$. The merger rate shape exhibits two distinct regimes, which we can explain by examining the upper limit of integration in \cref{eq:averagevrel}. We visualize the function governing this limit in \cref{fig:MR2}, along with the value of the escape velocity $v_\mrm{esc}(r)$.

When the merger condition is suppressed by the relative velocity between solitons and the integration upper limit is dominated by $v^*_{\text{rel}}$, the velocity saturating \cref{eq:mergercond}, the effective cross section in \cref{sigmaeff} is dominated by the geometrical cross section, such that $\sigma_\mrm{eff} \sim 4 \pi R_\mrm{sol}^2$. Thus, $\langle \sigma_{\text{eff}}\,v_{\text{rel}}  \rangle_{\text{merg}} \propto v^{*\,4}_{\text{rel}} R^2_{\text{sol}} \propto (M_{\text{sol}} m)^4 (M^{-1}_{\text{sol}} m^{-2})^2 \sim M_\mrm{sol}^2$. This cancels  with the factor of $M^{-2}_{\text{sol}}$ from the soliton number density in \cref{eq:MR}, leading to the merger rate being independent of the soliton and dark photon masses.  This regime occurs for $M_{\text{sol}}m \lesssim 10^{-13}\, M_{\odot}\,\text{eV}$. On the other hand, when the escape velocity dominates the limit, the effective cross section is instead dominated by the gravitational focusing term. 
Thus, $\langle \sigma_{\text{eff}}\,v_{\text{rel}}  \rangle_{\text{merg}} \propto m^{-2}$. Accounting for the factor of $M^{-2}_{\text{sol}}$ from the soliton number density, this leads to $\Gamma_{\text{merg}} \propto M^{-2}_{\text{sol}}m^{-2}$. This regime occurs for $M_{\text{sol}}m \gtrsim 3\times 10^{-13}\, M_{\odot}\,\text{eV}$.\footnote{The same behavior was explained in the case of scalar solitons in Ref.~\cite{Hertzberg:2020dbk}, but within a different parameter space related to axion DM.}

\section{Results for All Initial Profile Indices $\gamma$}
\label{sec:app-all_gamma}

\begin{figure*}

    \centering
    \begin{subfigure}{0.495\textwidth}
    \includegraphics[width=0.93\textwidth]{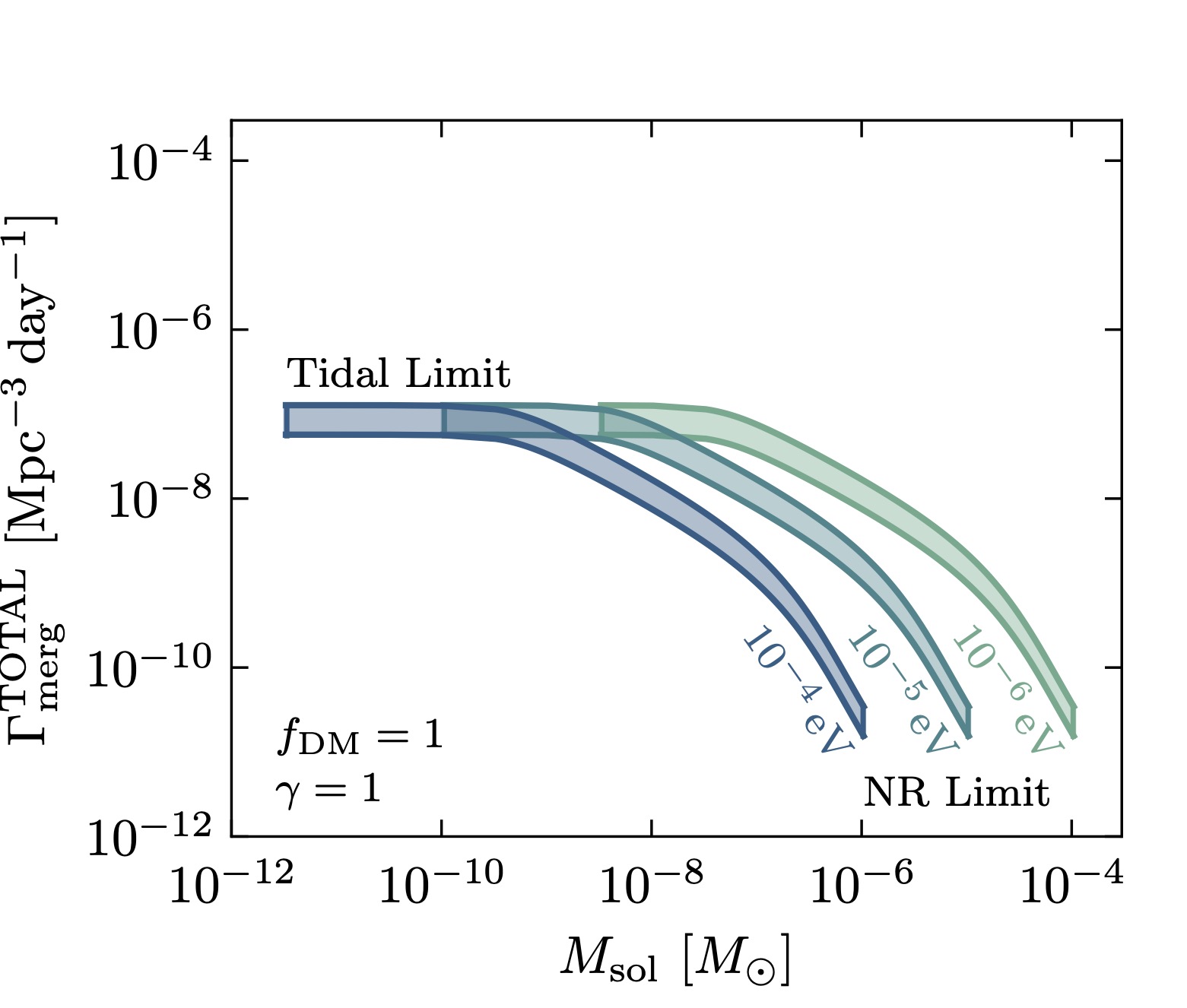}
    \end{subfigure}
    \begin{subfigure}{0.495\textwidth}
    \includegraphics[width=0.93\textwidth]{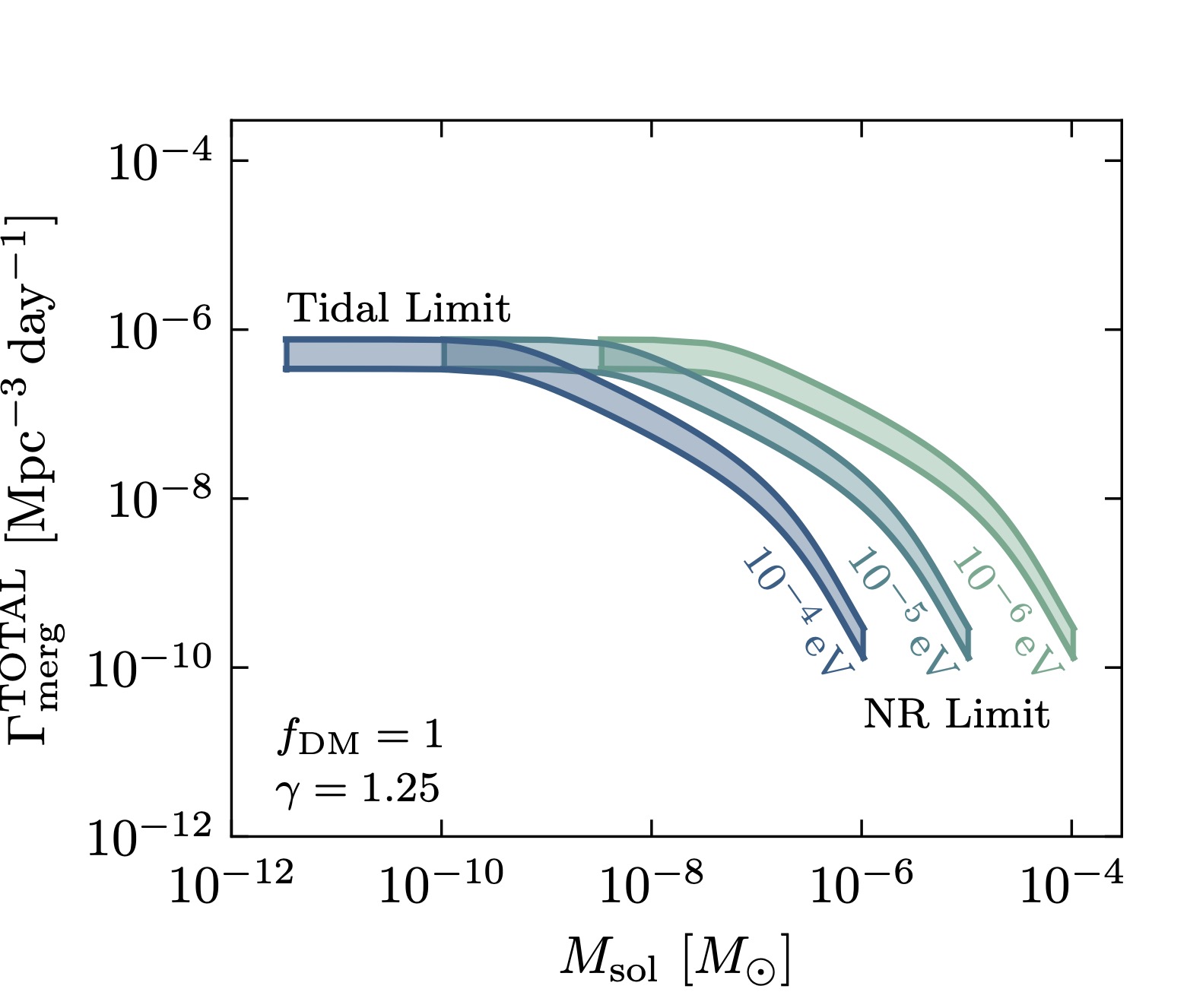}
    \end{subfigure}
    \begin{subfigure}{0.495\textwidth}
    \includegraphics[width=0.93\textwidth]{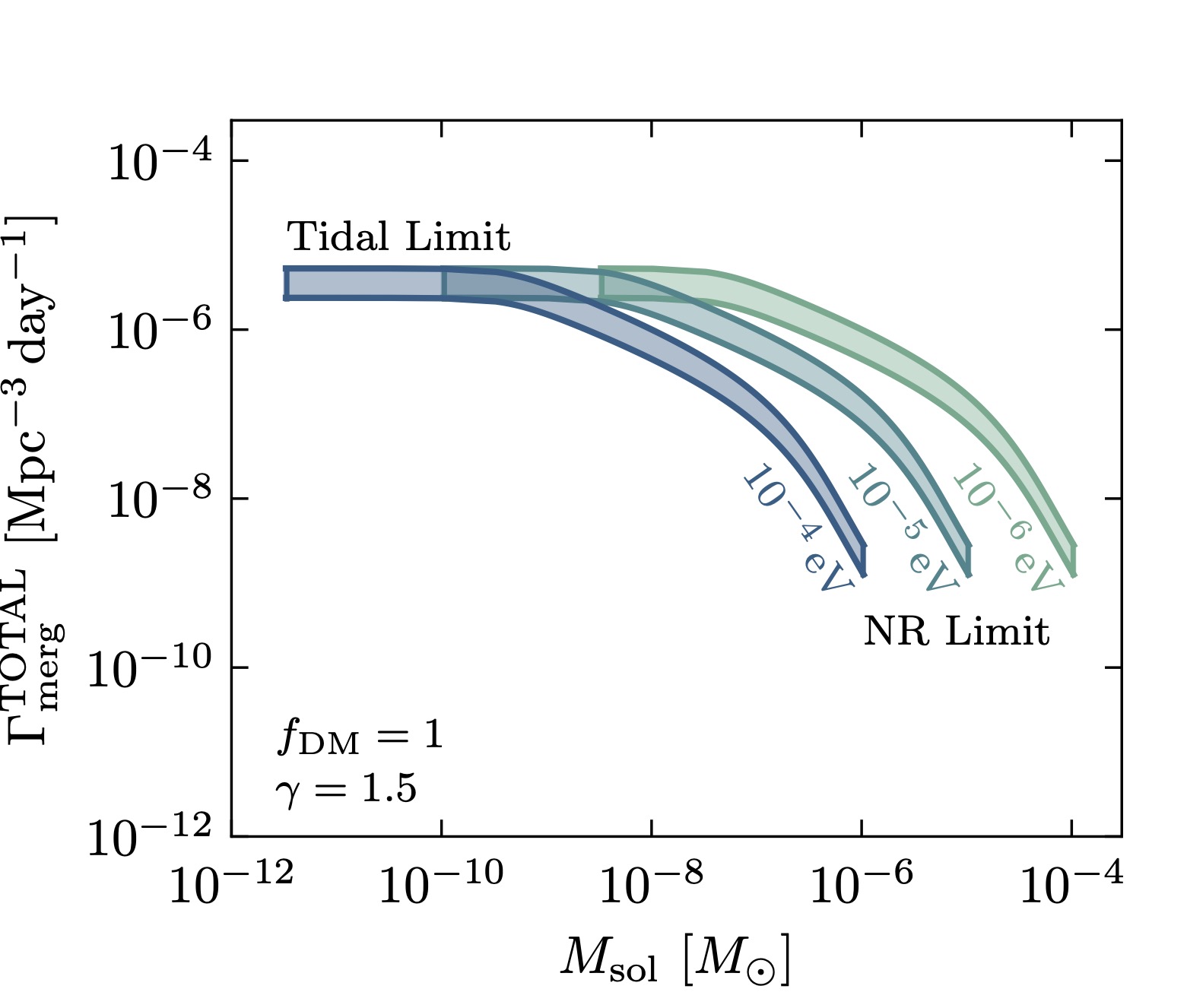}
    \end{subfigure}
    \begin{subfigure}{0.495\textwidth}
    \includegraphics[width=0.93\textwidth]{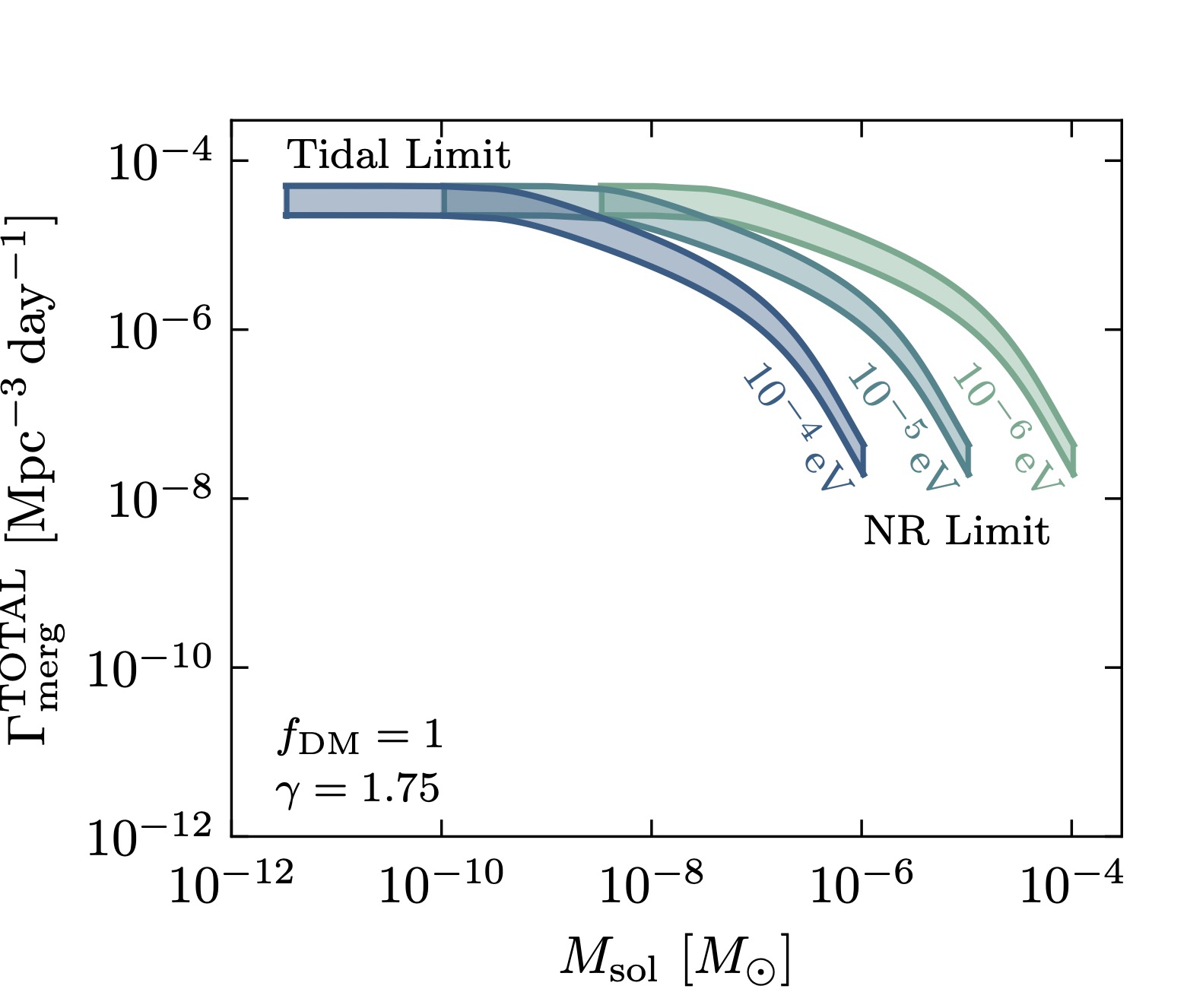}
    \end{subfigure}
    \caption{\justifying The total merger rate of vector solitons in DM spikes per unit volume $\Gamma_\mrm{merg}^\mrm{TOTAL}$ integrated over supermassive black holes with masses in the range $10^6\, M_{\odot} \leq M_{\text{SMBH}}\leq 8 \times 10^8\, M_{\odot}$  from $z=0$ to $z=4$ in terms of the soliton mass $M_\mrm{sol}$. Results are shown for all initial dark matter profile indices $\gamma$ we have considered, $\gamma \in \{1, 1.25, 1.5, 1.75\}$. Each colored band assumes a particular value for the dark photon mass, which we show for the range $m \in \{10^{-6}\,\mrm{eV}, 10^{-5}\,\mrm{eV}, 10^{-4}\,\mrm{eV}\}$.  We assume that dark photon solitons compose all of the dark matter, such that $f_\mrm{DM}=1$, and we take $\gamma=1$ for the initial dark matter profile, including the uncertainties in the supermassive black hole mass functions shown in \cref{fig:PhiSMBH}. The lower bound on the soliton mass is governed by the limit in which they can be tidally disrupted by supermassive black holes (\cref{eq:msol_tidal}), and the upper bound is dictated by remaining in the non-relativistic (NR) limit (\cref{eq:nonrelbound}). See also \cref{fig:MR}.}
    \label{fig:MR_app}
\end{figure*}

\begin{figure*}
    \centering
    \begin{subfigure}{0.495\textwidth}
    \includegraphics[width=0.93\textwidth]{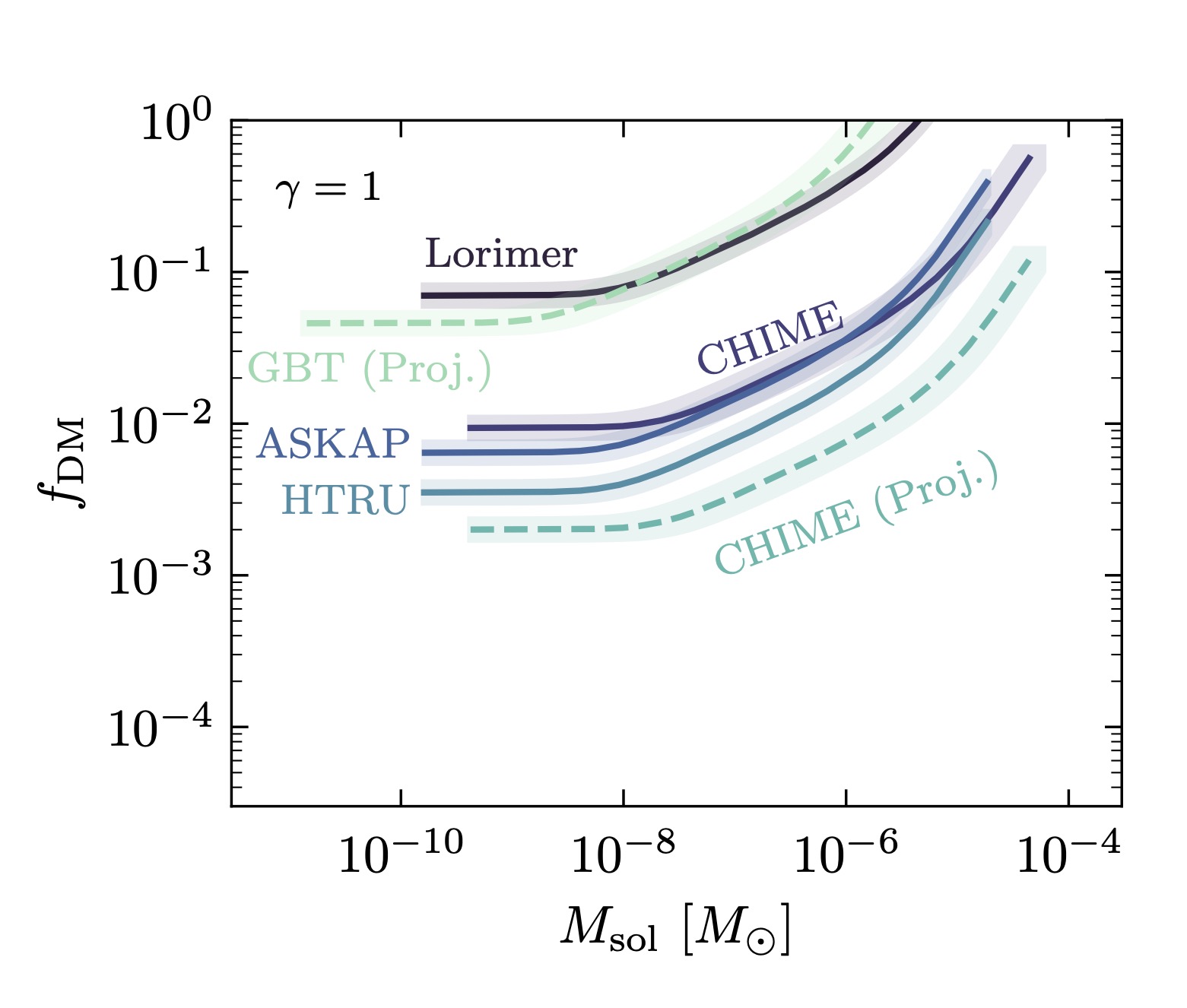}
    \end{subfigure}
    \begin{subfigure}{0.495\textwidth}
    \includegraphics[width=0.93\textwidth]{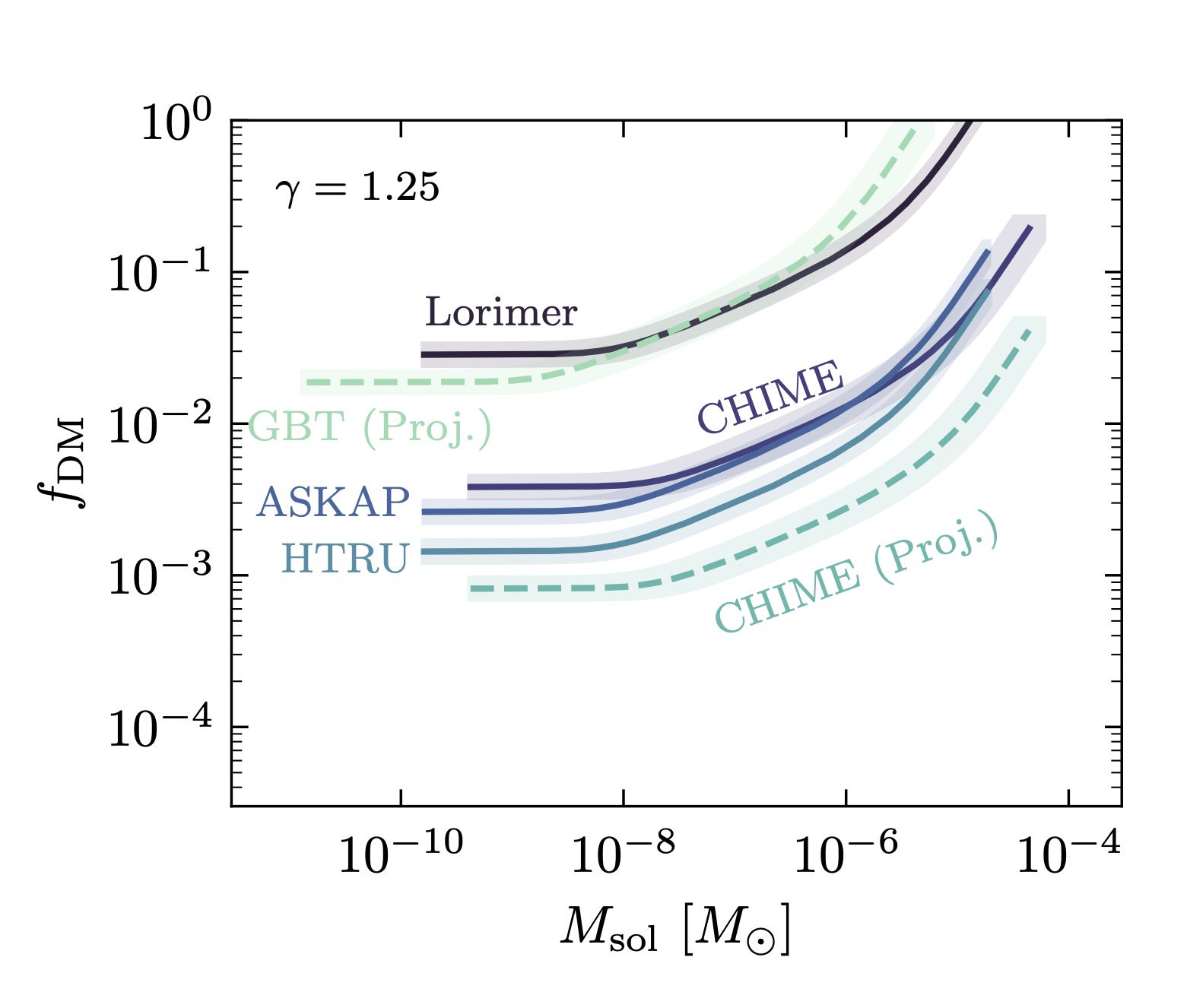}
    \end{subfigure}
    \begin{subfigure}{0.495\textwidth}
    \includegraphics[width=0.93\textwidth]{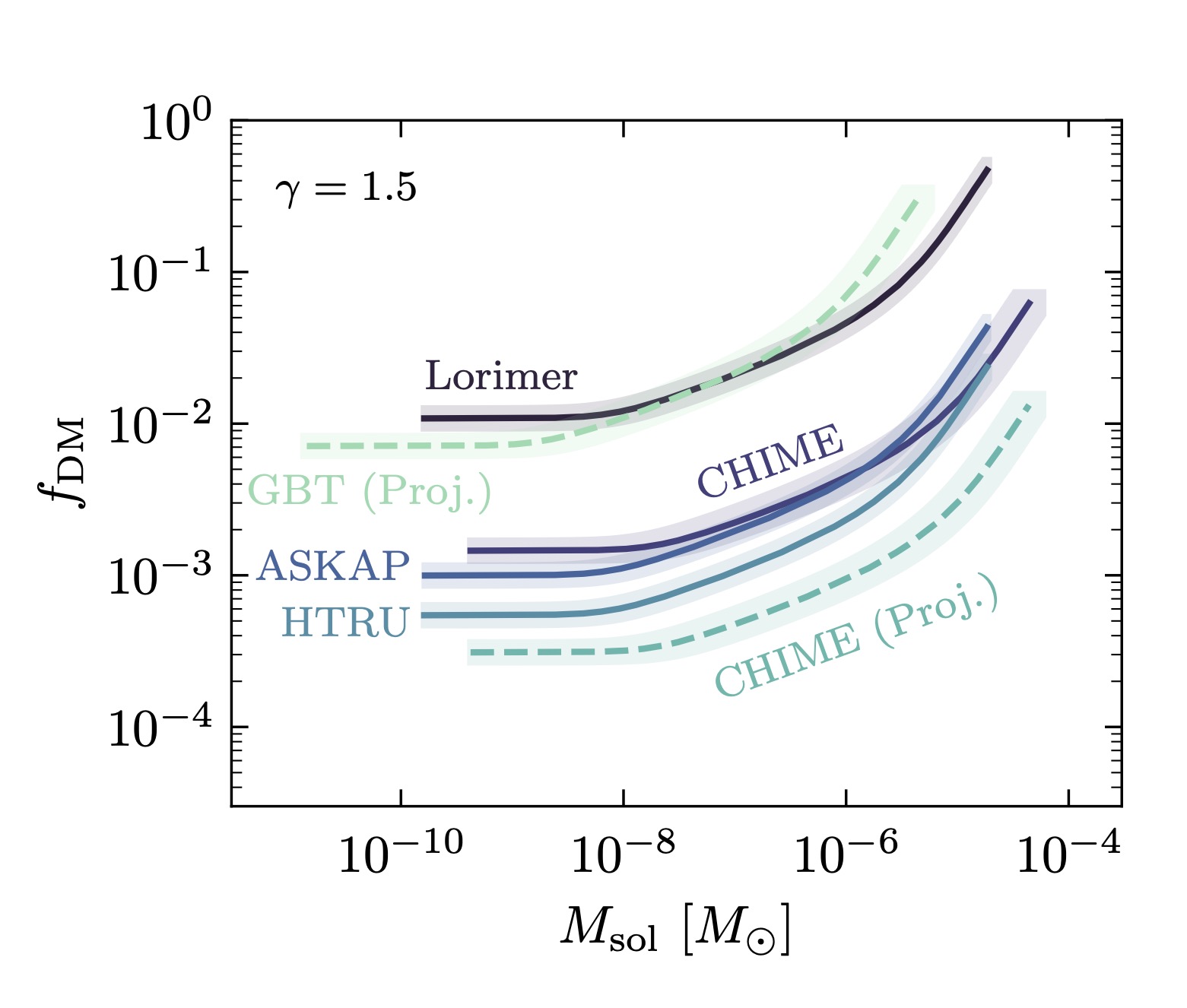}
    \end{subfigure}
    \begin{subfigure}{0.495\textwidth}
    \includegraphics[width=0.93\textwidth]{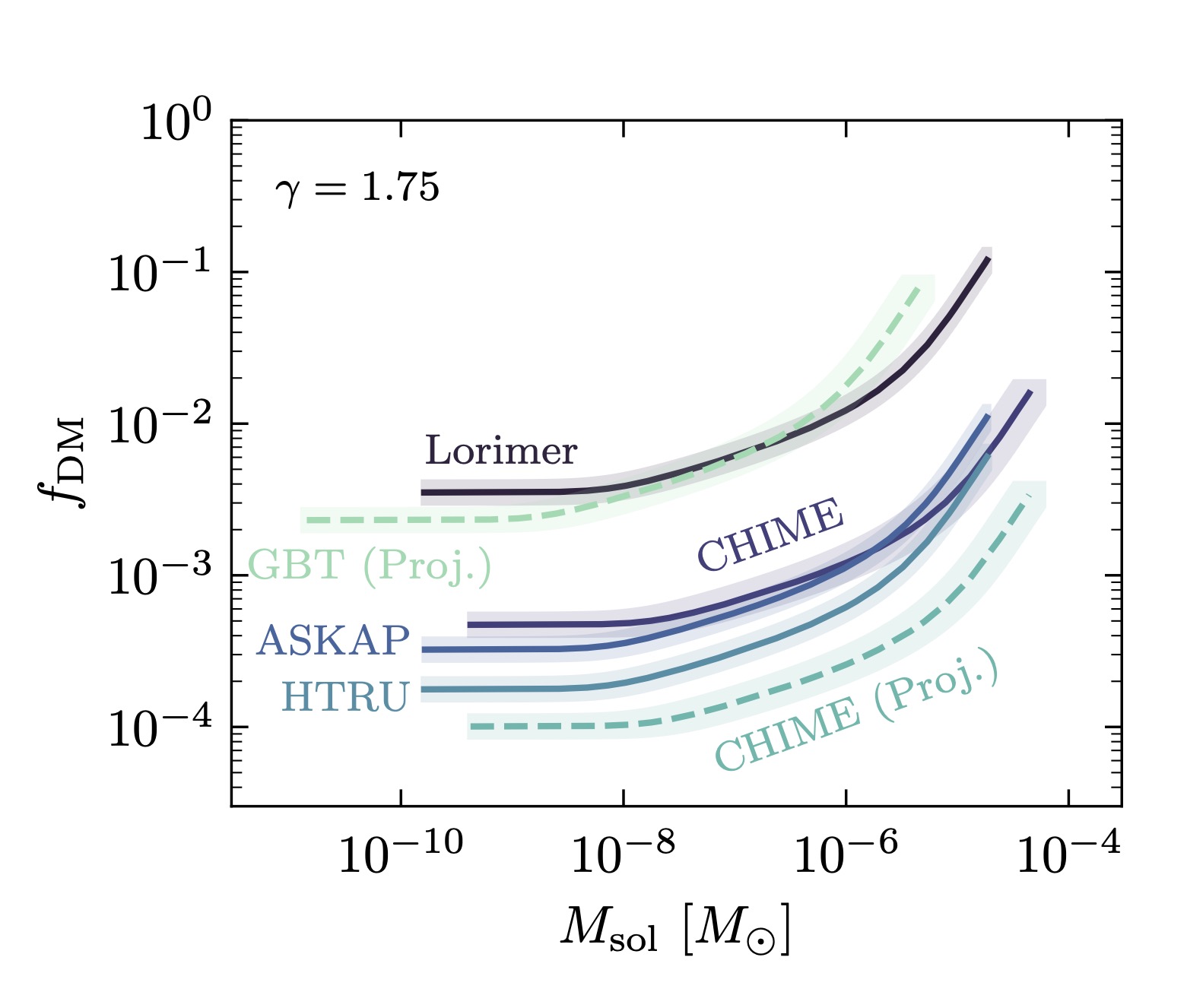}
    \end{subfigure}
    \caption{\justifying The $95\%$ confidence level upper limits on the fraction of dark matter that vector solitons can compose $f_\mrm{DM}$ with soliton mass $M_\mrm{sol}$. Limits and projections are shown for the telescopes considered in \cref{tab:telescopes-properties}. The bands correspond to the where each limit can lie as a result of the uncertainty in the merger rate, shown in \cref{fig:MR}, and the range of frequencies probed by each telescope. Each band commences at the minimum soliton mass required for tidal stability around supermassive black holes (\cref{eq:msol_tidal}) and terminates when the non-relativistic (NR) approximation becomes invalid (\cref{eq:nonrelbound}). The solid lines show the geometric mean of the limits at the boundaries of each band, which we take to be the best estimator of our limits. Results are shown for all initial dark matter profile indices $\gamma$ we have considered, $\gamma \in \{1, 1.25, 1.5, 1.75\}$. See also \cref{fig:fdm_lim}.}
    \label{fig:fdm_lim_app}
\end{figure*}

We present the main results of our work for all of the initial dark matter profile indices we have considered: $\gamma \in \{1, 1.25, 1.5, 1.75\}$. We show the total soliton merger rate in the vicinity of the spiky dark matter density profiles formed by central galactic SMBHs in \cref{fig:MR_app}. This is explained in \cref{sec:merger_rate}. We also show the $95\%$ confidence level limits on the dark matter fraction that vector solitons can compose in \cref{fig:fdm_lim_app}. The details of this calculation are given in \cref{Sec:Constrain} and elaborated on throughout \cref{sec:param-res}.

\newpage

\clearpage\newpage

\section{Caveats and Assumptions}
\label{sec:assumptions}

\subsection{Relativistic Doppler Shift}
\label{subsec:doppler}

For the mergers occurring near SMBHs, there will be a Doppler broadening, resulting in a Doppler scale factor $\mathcal{D}$ for both the central frequency in \cref{eq:nu} and the signal bandwidth in \cref{eq:delta_nu}. Given a soliton traveling at a speed $\beta c$ away from our line of sight at an angle $\theta$ with respect to us, this scale factor is given by
\begin{equation}
\mathcal{D}(\beta, \theta) = \frac{1}{\gamma (1 + \beta \cos \theta)}\,.
\end{equation}
In the extreme cases, a given merger can occur either moving directly away from our line of sight ($\theta = 0$) or towards it ($\theta = \pi$). The range for the Doppler factor is thus
\begin{equation}
\mathcal{D} \in \left[\sqrt{\frac{1 + \beta}{1 - \beta}}, \sqrt{\frac{1 - \beta}{1 + \beta}}\right]\,.
\end{equation}
Moreover, given that we should have a uniform prior on $\cos \theta$ and that the merger events are rare, the expected Doppler factor is
\begin{equation}
\langle\mathcal{D}\rangle = \frac{1}{2} \int_{-1}^{1} \mathcal{D} \dd\!\cos\theta = \frac{1}{2\gamma \beta} \ln\left(\frac{1+\beta}{1-\beta}\right)\,,
\end{equation}
with the second moment similarly given by
\begin{equation}
\langle \mathcal{D}^2 \rangle  = \frac{1}{\gamma^2(\beta^2 - 1)}\,.
\end{equation}
This allows us to compute the variance of the distribution $\sigma_\mathcal{D}^2 = \langle\mathcal{D}^2\rangle - \langle\mathcal{D}\rangle^2$.

The question then becomes what should be taken for $\beta$. The radiation is emitted by merged solitons, and while we may readily predict the relative velocities of solitons pre-merger, the velocity of the merged soliton would depend on the details of the merger dynamics. As an estimate, however, we will conservatively take this velocity to be of order the relative velocities of solitons, such that $\beta \sim v_\mrm{rel} / c$. 

The distribution of relative velocities is well described by a 3D Maxwell-Boltzmann distribution, as per \cref{eq:Pvrel}. From this, we find that $\langle v_\mrm{rel}(r)\rangle \sim \sigma_\mrm{rel}(r) \sim \sigma_\mrm{DM}(r)$, with $\sigma_\mrm{DM}(r)$ given by \cref{eq:Jeq}. Since we take the minimum distance from the SMBH to be twice the Schwarzschild radius $r_\mrm{Sch} = (2 G_\mrm{N} M_\mrm{SMBH} / c^2)^{1/2}$, we have that 
\begin{equation}
\beta_\mrm{max} \sim \frac{\sigma_\mrm{DM}(r = 2 r_\mrm{Sch})}{c} = \frac{1}{2\sqrt{2}} \approx 0.4\,.
\end{equation}
Though the velocities are more accurately described in Ref.~\cite{Zhang:2025mdl}, this functions to give us an estimate of the Doppler shift. Inserting this into our Doppler expressions, we find
\begin{equation}
\mathcal{D} \in [0.69, 1.4], ~~\text{and} ~~ \langle \mathcal{D} \rangle \approx 0.98 ~~ \text{with} ~~ \sigma^2_\mathcal{D} \approx 0.06\,.
\end{equation}

Observationally, the key change in our analysis would arise from the shift in the central frequency of the radiation. From the above calculation, this shift can be as large as $\usim 30\%\text{--}40\%$ in the extreme cases. As $m = 2 \pi \nu$, this leads to an equivalent uncertainty in the `reconstructed' dark photon mass. However, we more realistically expect this factor to be sharply centered at around $\mathcal{D} \sim \langle \mathcal{D}\rangle \approx 1$. Our results are therefore not appreciably impacted by the Doppler effect.

\subsection{Radiation Escape from Galactic Centers}
\label{subsec:radiationescape}

A potential caveat to interpreting non-detections as `radio silence' is that the escape of
$\usim 10^2$--$10^3~{\rm MHz}$ radio-frequency radiation from the inner regions of galactic centers can be nontrivial. However, as we argue below, we do not expect this to impact our results, with only a mild relaxation of bounds in the worst case scenario.
At these radio frequencies, the dominant attenuation mechanism in ionized gas is typically free--free
(absorption) opacity, which can be expressed in terms of the emission measure
${\rm EM}\equiv \int n_e^2 \dd l$ as \cite{1967ApJ...147..471M, 1975A&A....39....1P}
\begin{equation}
\label{eq:tau_nu}
\tau_{\rm ff}(\nu)\simeq 3.3\times 10^{-7}\,
\left(\frac{T_e}{10^4\,{\rm K}}\right)^{-1.35}
\left(\frac{\nu}{1\,\rm GHz}\right)^{-2.1}
\left(\frac{\rm EM}{1\,\rm pc\,cm^{-6}}\right)\,,
\end{equation}
where $\tau_{\rm ff}(\nu)$ is the free–free optical depth at frequency $\nu$, and $T_e$ and $n_e$ are the temperature and number density of free electrons, respectively. A larger value for the optical depth means that more of the radiation is attenuated. Because ${\rm EM}$ and $T_e$ vary widely among galactic nuclei (from hot/dilute accretion flows
to warm/dense ionized streamers or star-forming environments), the emergent transmission
$\mathcal{T}(\nu)\sim e^{-\tau_{\rm ff}(\nu)}$ can range from negligible attenuation to strong suppression.

As an illustration of this, the hot/dilute case ($T_e\sim 10^7~{\rm K}$, ${\rm EM}\sim 10^{4}$--$10^{5}~{\rm pc\,cm^{-6}}$)
gives $\tau_{\rm ff}(0.1~{\rm GHz})\sim 10^{-5}$--$10^{-4}$, whereas the warm/dense case
($T_e\sim 10^4~{\rm K}$, ${\rm EM}\sim 10^{6}$--$10^{8}~{\rm pc\,cm^{-6}}$) yields
$\tau_{\rm ff}(0.1~{\rm GHz})\sim 10^{1}$--$10^{3}$ and can remain $\gtrsim \!\mathcal{O}(1)$ even near a GHz.
Such low-frequency absorption has long been discussed in the Galactic Center environment
\cite{1976MNRAS.177..319D, 1989ApJ...342..769P, 2004ApJ...601L..51N, 2004MNRAS.349L..25R}
and is also expected to shape the low-frequency spectra of gas-rich starbursts \cite{2013MNRAS.431.3003L}.
In addition, strong scattering in dense/turbulent nuclear plasma can broaden short-duration
transients at low frequencies \cite{Cordes:2002wz}, and induced scattering can affect
extremely bright coherent pulses in sufficiently dense plasma close to the source \cite{2008ApJ...682.1443L, Metzger:2019una}.

Moreover, hot/dilute and warm/dense regions can coexist and describe different phases of the nuclear interstellar medium. It is entirely plausible to have a hot RIAF (Radiatively Inefficient Accretion Flow) very close to the SMBH, plus warm ionized gas (streamers/H II) on $\usim \mrm{pc}$ scales, plus molecular/dusty structures farther out. Indeed, the Milky Way is a good example of this multi-phase situation: Sagittarius A$^*$ is modeled as a RIAF/LLAGN (Low-Luminosity Active Galactic Nucleus)-like source while the central parsecs also host ionized structures (Saggitarius A West) that can provide appreciable free–free absorption at low frequencies\,\cite{2018MNRAS.478.3544R, 2004MNRAS.349L..25R}. The relevant phase regarding our situation would depend strongly on the line of sight and nuclear conditions.

Combining all effects, any attenuation in the host nucleus reduces the effective detection efficiency by a factor
\begin{equation}
\mathcal{T}(\nu) \equiv e^{-\tau_{\text{ff}}(\nu)}\times (\text{scattering/selection effects})\,.
\end{equation}
These effects only serve to decrease the expected number of detected events for a given intrinsic merger rate and, consequently, accounting for them would relax our bounds on $(f_{\text{DM}},g,m)$. However, the extent to which they would relax depends on the details of interstellar medium and radiation frequency. 

Concretely, the fluence observed by a telescope can be written as
\begin{equation}
\label{eq:sb_atten}
    \mathcal{F}_\mrm{obs}(\nu) = \mathcal{T}(\nu)\mathcal{F}_0\,,
\end{equation}
where $\mathcal{F}_0$ is the unattenuated value, as given by \cref{eq:fluence}. If the attenuation is strong enough, the `observed' fluence can fall below the observational threshold value for a given telescope $\mathcal{F}_\mrm{thresh}$. In practice, free–free absorption dominates over scattering/selection whenever the medium is optically thick at the observing frequency, i.e.~when $\tau_\text{ff}(\nu) \gtrsim 1$. This will be true in our analysis below, so we neglect scattering and selection effects. We can then solve for that free--free opacity above which a signal can be attenuated below the observational sensitivity of a telescope. Using \cref{eq:fluence}, we find
\begin{equation}
    \tau_\mrm{ff}^\mrm{max} \equiv \ln\left[\frac{\mathcal{F}_0}{\mathcal{F}_\mrm{thresh}}\right] \simeq \ln\left[\left( \frac{10^{-6} \, \mathrm{eV}}{m} \right)^{2}\left( \frac{1 \, \mathrm{Mpc}}{D_\mrm{L}} \right)^{2} \left(\frac{2 \times 10^{16}\,\mrm{Jy\,ms}}{\mathcal{F}_\mrm{thresh}}\right)\right]\,.
\end{equation}
The values for the fluence thresholds range from $\mathcal{F}_\mrm{thresh}\sim 0.1\,\mrm{Jy\,ms}\text{--}10\,\mrm{Jy\,ms}$~\cite{2016MNRAS.455.2207R,Gajjar:2018bth,CHIMEFRB:2018mlh,2018Natur.562..386S}. In fact, for our approximation, only the final term in parentheses matters due to the logarithmic dependence of $\tau_\mrm{ff}^\mrm{max}$. For the masses and distances we consider, we find the smallest and therefore most problematic value for this opacity to be $\tau^\mrm{max}_\mrm{ff} \approx 10$, occurring when the distance over which the radiation must travel is maximized at $D_\mrm{L} \approx 7332\,\mrm{Mpc}$ (occurring at our largest considered redshift value of $z = 4$). Then, assuming the worst possible case scenario where this radiation must travel through a warm/dense region, we get using \cref{eq:tau_nu} that frequencies $\nu \lesssim 0.15\,\mrm{GHz}$ for $\mrm{EM} \sim 10^6\,\mrm{pc\,cm^{-6}}$ or $\nu \lesssim 1.5\,\mrm{GHz}$ for $\mrm{EM} \sim 10^8\,\mrm{pc\,cm^{-6}}$ are attenuated below typical fluence threshold values.

The upper frequency bound of the $\mrm{EM} \sim 10^6\,\mrm{pc\,cm^{-6}}$ scenario is smaller than the lowest telescope frequency we consider of $\nu = 0.4\,\mrm{GHz}$ for CHIME (cf.~\cref{tab:telescopes-properties}). In this case, we therefore do not expect any impact on our results. On the other hand, for  $\mrm{EM} \sim 10^8\,\mrm{pc\,cm^{-6}}$, the upper bound of $\nu \lesssim 1.5\,\mrm{GHz}$ appears to render all telescopes except for GBT blind to our considered signals, such that our limits would only apply in the case that $\usim1\,\mrm{GHz}$ can escape from galactic centers. However, even in this case,  we do not expect appreciable changes to our results due to radiation attenuation within the interstellar medium.

Specifically, \cref{fig:PhiSMBH} shows the supermassive black hole mass function in terms of the black hole mass and redshifts. This population includes many quiescent/dormant systems (especially at low redshift, much of the population is not in a luminous/AGN phase), for which the inner environment can be gas-poor or dominated by hot, dilute plasma rather than a dense H II-like absorber\,\cite{Tucci:2016tyc}. In addition, the clumpy structure of typical warm/dense optical thick regions in the galactic inner regions allows some sightlines to intersect dense streamers (high EM → large $\tau_{\text{ff}}$), while others mostly traverse lower-EM hot/dilute gas. Therefore, due to the nature of the SMBH population that we considered and the discontinuous inner structure of typical optically thick inner galactic regions, we expect that the relaxing effect over our bounds in the worst-case scenario is still mild and terrestrial telescopes will not be blind to our signals.

\subsection{Beyond the instantaneous burst approximation}
\label{subsec:instburst}

We are working in the `instantaneous burst' approximation, as explained in Ref.~\cite{Amin:2023imi}, where the resonance phenomenon is triggered once the merger has settled down into a new solitonic ground-state solution of the multi-component SP system of differential equations. Indeed, \cref{eq:flux_density} considers the maximum amount of evaporated mass in the minimum duration time. Considering that the typical scale for resonance is much smaller than the vector soliton scale, $\lambda \sim 1/m \ll R$, and the merger process takes a certain finite time, we could expect to have a successively smaller consecutive burst than a brighter unique burst\,\cite{Amin:2022pzv, Hertzberg:2020dbk}. One way to estimate this effect is to consider the typical time for a solitonic merger, which has been shown to be of the order of $\tau_{\text{eff}} \sim 1\,\text{yr}$, so that the spectral flux density in \cref{eq:flux_density} would take the form $\mathcal{S}_B = \mathcal{S}_{B,0} (\tau/\tau_{\text{eff}}) \sim 10^{-12} \times \mathcal{S}_{B,0}$\,\cite{Hertzberg:2020dbk}. Even in this case, the typical spectral flux density is many orders of magnitude larger than the spectral flux density thresholds associated with the radio telescope surveys that we are considering, $S_{B, \mrm{thresh}}\sim 0.1\,\mrm{Jy}\text{--}10\,\mrm{Jy}$~\cite{2016MNRAS.455.2207R,Gajjar:2018bth,CHIMEFRB:2018mlh,2018Natur.562..386S}, so that our radio silence framework is not jeopardized.

\subsection{Non-monochromatic Vector Soliton Mass Function}
\label{subsec:solitonnonmono}

The soliton mass function is expected to depend sensitively on the formation mechanism. For instance, in Ref.~\cite{Gorghetto:2022sue}, the number density of solitons produced via isocurvature density collapse is sharply peaked around a characteristic mass scale, whereas solitons formed through gravitational relaxation in the kinetic regime exhibit a mass distribution that depends on the properties of the host halo~\cite{Levkov:2018kau}. In this work, we have aimed to remain as model-independent as possible and therefore have focused on the fraction of solitons at a given mass. To achieve this, we have assumed a simple monochromatic soliton mass function centered around the soliton critical mass $M_c$.

Nonetheless, we can provide an idea of how our calculation may change when accounting for this. Take an arbitrary initial mass function immediately after soliton formation of $\phi(z_0, M_\mrm{sol})$, where $z_0$ is the redshift at which the solitons form. After formation, all supercritical mass solitons will undergo parametric resonance, ending with a mass just below the critical mass threshold $M_c$. Then the late-time soliton population, described by $\phi(z\ll z_0, M_\mrm{sol})$, will contain only subcritical solitons (except for some supercritical solitons formed around $z\approx0$ due to major merger events). In our scenario, these solitons end up populating the inner galactic regions around SMBHs, and soliton merger events produce the supercritical soliton masses that generate our desired signal via the parametric resonance phenomenon. 

Considering the merger of two solitons with masses $M_\text{sol,1}\approx M_\text{sol,2} \approx \widetilde M_\text{sol} $, the golden rule for mergers indicates that the condition for subsequent resonance for the resultant soliton reads as  $\widetilde M_\text{sol} \gtrsim (5/7)M_c$~\cite{Hertzberg:2020dbk, Schwabe:2016rze}. Thus, we may estimate the total fraction of solitons that actually participate in the resonance phenomenon in galactic inner regions, $f_\text{burst}$, as 
\begin{align}
f_\text{burst}&\sim\int_{5M_c/7}^{M_c}\phi(z_0,M_\mrm{sol})\dd M_\mrm{sol} + \int_{M_c}^{\infty} \phi(z_0,M_\mrm{sol})\dd M_\mrm{sol}\,.
\end{align}
The first integral corresponds to those subcritical solitons that may undergo major mergers by virtue of being close to $M_c$ at the time of their formation, whereas the second integral corresponds to those supercritical solitons that parametrically resonate down to masses around the critical mass at late times. This heuristic argument allows us to predict this fraction without explicitly computing the mass function $\phi(z, M_\mrm{sol})$, which would require a more technical treatment. For instance, our argument neglects the fact that subcritical solitons can merge into the mass region that would permit them to merge around $z\approx 0$ and contribute to our signal. Then, since the upper bound for the allowed DM fraction in solitons scales in \cref{eq:fDMlimit} as $[\Gamma_\text{merg}^\text{TOTAL}(f_\text{DM}=1)]^{-1/2}$, our constraint should weaken by a factor $f^{-1}_\text{burst}$.

For a concrete example, take an initial power-law mass function of the form
\begin{align}
\phi(z_0,M_\mrm{sol}) \equiv
\begin{cases}
 \mathcal{N} \left(M_\text{sol}/{M} \right)^{-\alpha} & \text{if}
\,\, M_\text{0} \leq M_\text{sol} \leq M_\text{1}\,,\\ 
0 & \text{else}\,,
\end{cases}
\end{align}
with $M_\text{0}$ and $M_\text{1}$ the minimum and maximum soliton masses able to be formed by the production mechanism, and the mass $M$ an arbitrary mass scale. The normalization constant $\mathcal{N}$ ensures that $\phi$ integrates to $1$ over all possible soliton masses, and $0 \leq \alpha < 1$ is a power-law index. Such a power-law mass function is well-motivated by \cite{Escudero:2023vgv}. For $M_0 < (5/7)M_c$ and $M_0 < M_c < M_1$, we then find that
\begin{equation}
    f_\mrm{burst} \simeq \frac{1 - (5/7)^{1-\alpha}(M_c/M_1)^{1-\alpha}}{1 - (M_0/M_1)^{1-\alpha}}\,.
\end{equation}
From this, we can retrieve two extreme cases. If $M_0 \ll M_1$ and $M_c \ll M_1$, then $f_\mrm{burst} \approx 1$ and our limits should be unchanged. On the other hand, if instead $M_c \rightarrow M_1$ and $\alpha \rightarrow 1$, then $f_\mrm{burst} \rightarrow 0$ and we expect our limits to lift off considerably. 
This example therefore serves to illustrate how sensitive our calculation is to the choice of production mechanism. To avoid assumptions on this, we have chosen a mass function that allows us to easily account for arbitrary late-time mass functions as per general the logic above.

\null\vfill

\clearpage

\bibliographystyle{JHEP.bst}
\bibliography{biblio.bib}
\end{document}